\documentclass[11pt]{article}

\usepackage{amsfonts}
\usepackage{amsmath}
\usepackage{esint}
\usepackage[margin=1in]{geometry}
\usepackage{graphicx}
\usepackage{epstopdf}
\usepackage{amssymb}
\usepackage{cite}
\usepackage{hyperref}

\usepackage{caption}
\usepackage{color}

\usepackage{multicol}
\numberwithin{equation}{section}

\usepackage[title]{appendix}

\newcommand{\ds} {\displaystyle}

\newcommand{\R}{{\mathbb R}}

\title{Analyzing the Effects of Observation Function Selection in Ensemble Kalman Filtering for Epidemic Models} %%%%%%%%%
\author{Leah Mitchell, Andrea Arnold*}
\date{}

\begin{document}
\maketitle

% Author affiliations
\small 
\centerline{Department of Mathematical Sciences, Worcester Polytechnic Institute, Worcester, MA, USA}
\vspace{.2cm}

\centerline{*corresponding author: anarnold@wpi.edu}

\normalsize

\bigskip

\begin{abstract}
The Ensemble Kalman Filter (EnKF) is a popular sequential data assimilation method that has been increasingly used for parameter estimation and forecast prediction in epidemiological studies.  The observation function plays a critical role in the EnKF framework, connecting the unknown system variables with the observed data.  Key differences in observed data and modeling assumptions have led to the use of different observation functions in the epidemic modeling literature.  In this work, we present a novel computational analysis demonstrating the effects of observation function selection when using the EnKF for state and parameter estimation in this setting.  In examining the use of four epidemiologically-inspired observation functions of different forms in connection with the classic Susceptible-Infectious-Recovered (SIR) model, we show how incorrect observation modeling assumptions (i.e., fitting incidence data with a prevalence model, or neglecting under-reporting) can lead to inaccurate filtering estimates and forecast predictions.  Results demonstrate the importance of choosing an observation function that well interprets the available data on the corresponding EnKF estimates in several filtering scenarios, including state estimation with known parameters, and combined state and parameter estimation with both constant and time-varying parameters.  Numerical experiments further illustrate how modifying the observation noise covariance matrix in the filter can help to account for uncertainty in the observation function in certain cases. \\

\end{abstract}

\bigskip
\bigskip

\noindent \textbf{Note:} This is the accepted manuscript of an article published in \textit{Mathematical Biosciences}. The published journal article is available online at \url{https://doi.org/10.1016/j.mbs.2021.108655}.\\[0.25cm]

\noindent To cite, please use: L. Mitchell and A. Arnold (2021) Analyzing the effects of observation function selection in ensemble Kalman filtering for epidemic models. \textit{Mathematical Biosciences}, 339, 108655. doi: 10.1016/j.mbs.2021.108655

\mbox{}
\vfill

\noindent \line(1,0){155}\\

\noindent \textcopyright\; 2021. This manuscript version is made available under the CC BY-NC-ND 4.0 license (\url{https://creativecommons.org/licenses/by-nc-nd/4.0/}).

\newpage

%%%%%%%%%%%%%%%%%%%%%%%%%%%%%%%%%%%%%%%%%%%%%%%%%%%
\section{Introduction}

Ensemble Kalman filtering is a popular data assimilation approach that has been increasingly used to estimate unknown system states and parameters in a variety of real-world problems with time series data \cite{evensen, burgers, enkftheory, Katzfuss_EnKF}.  Compared to other filtering algorithms, ensemble Kalman filters avoid particle degeneracy issues associated with resampling and have shown to be computationally feasible for high-dimensional problems \cite{Katzfuss_EnKF, Fearnhead2018}.  While these methods are commonly used in applications to weather prediction \cite{EnKFreview2016, weatherenkf, VarEnKF2017} and guidance, navigation, and control \cite{aerodynamic, attitudes, EnKFgnc2017}, ensemble Kalman-type filters have recently been utilized for parameter estimation and forecast prediction in a variety of epidemiological studies \cite{ScienceCovid2020, Narula2016, FFT2012, comparison, HFMD2019, HBV2018, arnold2018, Engbert2021}.

The filtering process comprises a two-step sequential updating scheme of forward prediction via a prescribed model and correction with the available data.  The inverse problem of estimating the system unknowns therefore involves defining two key components: the forward model and the observation model.  The forward model describes the dynamics of the system, predicting how the model states are propagated forward in time.  In this work, we consider ordinary differential equation (ODE) models of the form
\begin{equation}\label{eq:forward_model}
\frac{dx}{dt} = f(t,x,\theta), \hspace{.2in} x(0)=x_{0}
\end{equation}
where $f:\R\times\R^d\times\R^p\rightarrow\R^d$ is the righthand-side function defining the dynamics, $x=x(t)\in\R^d$ is a vector representing the states of the system, $\theta\in\R^p$ is the parameter vector, and $x_{0}\in\R^d$ is the initial condition.  
The observation model relates the data set back to the systems inputs, assuming discrete observations of the form
\begin{equation}\label{eq:obs_model}
y_{j}=g(x_{j},\theta) + e_{j}, \hspace{.2in} j=1,\dots,T
\end{equation}
where $y_{j}\in\R^m$ is an observation at time $t_j$, $g:\R^d\times\R^p\rightarrow\R^m$, $m\leq d$, is the observation function relating the model states $x_{j}=x(t_j)$ and parameters $\theta$, and $e_{j}\in\R^m$ is the observation error. 

While ensemble Kalman filtering is a powerful tool in state and parameter estimation, its performance depends not only on the choice of the forward model \eqref{eq:forward_model} driving the system dynamics, but also on the function $g$ in \eqref{eq:obs_model} used in modeling the observed data -- this function plays a vital role in well connecting the available data with the system variables.  Specifically, the observation function is utilized in the analysis step of the filter, when observed data are compared to model observation predictions computed using $g$.  Intuitively, it follows that using an observation that well represents the data should lead to correspondingly accurate estimates of the system unknowns.  However, it is not immediately clear how using a less accurate observation function affects filter performance, in particular as the filtering problems considered become more and more complex (e.g., estimating states with known parameters vs. jointly estimating unknown states and parameters).  Knowledge of this becomes especially important in applications where the best observation function is not immediately clear, or when modeling assumptions are necessary to simplify the situation compared to the actuality of data collection (i.e., when the bridge between the real data and the model input is more complex than that of the simplifying assumptions). 

In the setting of epidemiology, while significant effort has been put towards developing useful forward models for a variety of infectious diseases, often less emphasis is placed on the choice of the observation function representing the data \cite{Bolker2020}.  Forward models range from general compartment models \cite{greenbook, yellowbook, DC, SISandSIR} to more complex, disease-specific models describing, e.g., the spread of influenza \cite{comparison,pandemic,seasonality}, the AIDS epidemic \cite{greenbook,HIVAIDS}, and the novel coronavirus (COVID-19) \cite{ScienceCovid2020, FrontMed2020, DCcov2020, Calafiore2020, WangCov2020}.  These models are often variations of the well-known Kermack-McKendrick model \cite{KermackSIR} and are otherwise referred to as Susceptible-Infectious-Recovered (SIR) models, with compartments representing susceptible, infectious, and recovered portions of a population.  While SIR-type models are commonly used in governing the forward dynamics of epidemiological systems, a variety of different observation functions have been used in the literature to model the available data \cite{capaldi, arnold2018, Osthus2017, Calafiore2020, Marinov2014, Brookmeyer1990, Capistran2012, Pan2014, Yaari2018, Engbert2021}.  

\begin{figure}[!t]
 \centering
\includegraphics[width=\linewidth]{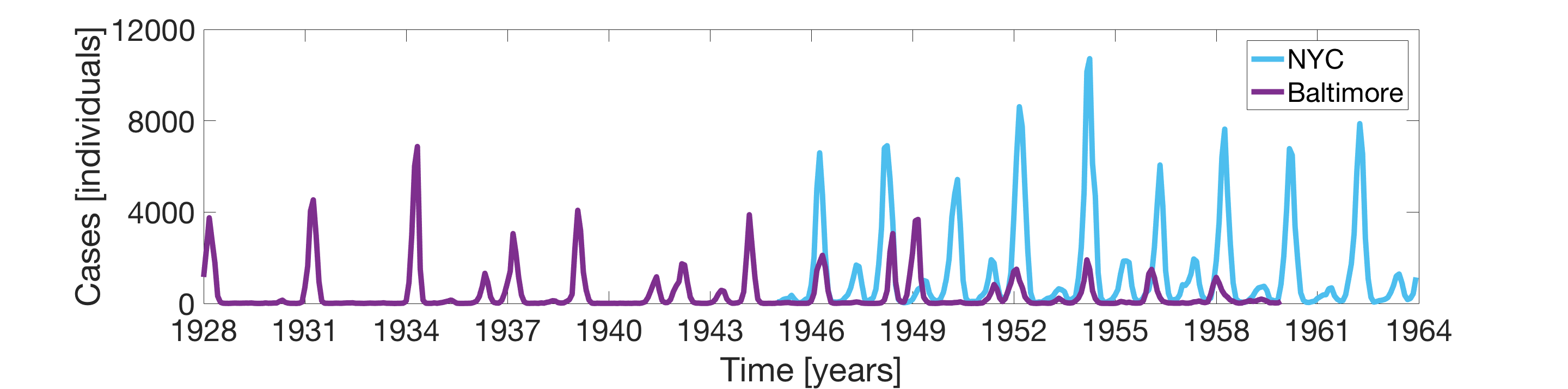}
\caption{Example data sets collected during measles outbreaks in the United States during the pre-vaccine era.  The purple line shows the monthly reported measles cases in Baltimore, MD, from 1928 to 1960; and the blue line shows the monthly reported cases in New York City, NY, from 1945 to 1964.  The data sets were obtained from an online infectious disease database (found at \url{https://ms.mcmaster.ca/~bolker/measdata.html}), as in \cite{arnold2018}.  
}
\label{fig:measles}
\end{figure}

Epidemiological data can vary based on whether the data report the number of newly infected individuals (incidence data) or the number of currently infected individuals (prevalence data) \cite{Bjornstad}, and whether the available data well represent the actual number of cases or if cases are significantly under-reported \cite{Gunning2014}.  Data may also differ by disease type, geographical spread, and frequency of reported cases (daily, weekly, monthly, etc.) \cite{yellowbook,greenbook,DC}.  For example, Figure \ref{fig:measles} shows reported measles incidence data from two cities in the United States with different population sizes, New York City (NYC) and Baltimore, MD, during the pre-vaccine era \cite{London}.  While each data set was collected on a monthly basis, there are clear differences in the magnitude and frequency of the outbreaks in each city: Baltimore starts with infrequent outbreaks of larger magnitude in the early years but tends towards smaller, more frequent outbreaks in the later years; on the other hand, NYC has a more regular bi-annual peaking pattern, with the larger outbreaks being typically much larger in magnitude than those of Baltimore.  The magnitudes of the outbreaks are further tied to the reporting probability of cases for each location, which is estimated to be around 1 in every 8 cases reported in NYC and 1 in every 3 or 4 cases reported in Baltimore during this time \cite{London}.

%%FIGURE

Different data sets and modeling assumptions have thereby led to the use of different observation functions in the epidemic modeling literature: This choice has varied from assuming direct measurements of the infectious population \cite{capaldi, Capistran2012, Marinov2014} to functions accounting for the cumulative number of cases over a given collection period \cite{arnold2018, Bolker2020, Yaari2018, Engbert2021}.  Some functions have also included a reporting probability to account for the under-reporting of cases \cite{arnold2018, Calafiore2020}.  Considering the different types of epidemiological data available, it may not be immediately clear which observation function is best suited for a given problem.  However, as recently shown in \cite{Bolker2020}, incorrect assumptions in modeling the observed data can lead to inaccurate and overly confident estimates of model parameters such as the basic reproduction number.

In this paper, we present a novel analysis which demonstrates how significantly the choice of observation function affects the performance of ensemble Kalman filtering in the setting of epidemic modeling.  In particular, we perform a simulation study to show how incorrect observation modeling assumptions (i.e., by mistaking incidence for prevalence data or neglecting under-reporting) can lead to inaccurate model state and parameter estimates, thereby diminishing the accuracy of forecast predictions.  We highlight the effects of using suboptimal observation functions in three different filtering scenarios of increasing complexity: state estimation with known parameters; combined state and constant parameter estimation; and combined state and time-varying parameter estimation.  We further demonstrate how inflating the observation noise covariance matrix can be utilized to help offset the effects in some filtering scenarios when under-reporting is not accounted for in the observation function.

The paper is organized as follows: Section~\ref{ch:kf} provides a review of Kalman filtering algorithms, first outlining the steps of the classic Kalman Filter and then describing three ensemble Kalman filtering variants for nonlinear systems.  Section~\ref{ch:SIR} details the SIR-type forward model and the four different observation functions used in this work to model the simulated epidemiological data.  Section~\ref{ch:results} contains the numerical results, which analyze the use of each observation function in the three different filtering scenarios described, as well as the effects of modifying the observation noise covariance.  Section \ref{ch:diss} presents a discussion of the results, and Section~\ref{ch:conclusion} gives a brief summary and conclusions.

%%%%%%%%%%%%%%%%%%%%%%%%%%%%%%%%%%%%%%%%%%%%%%%%%%
\section{Review of Kalman Filtering Methods}\label{ch:kf}

Named after Rudolf E. Kalman, an electrical engineer and mathematician who received the National Medal of Science for his work on the algorithm, the Kalman filter was famously used by NASA during the Apollo missions \cite{KFApollo,KFApolloIEEE} and in a variety of application areas since.  The algorithm was derived specifically for linear and Gaussian systems, utilizing the fact that Gaussian distributions remain Gaussian under linear transformation.  Various extensions of the Kalman filter have been derived to accommodate when these assumptions cannot be met.  Here we review the classic Kalman Filter (KF) and two of its nonlinear extensions, the Ensemble Kalman Filter (EnKF) for state estimation and the augmented EnKF for combined state and parameter estimation.

Kalman filtering comprises a two step process: predicting through the use of a forward model and correcting through the use of available data.  Under the Bayesian framework \cite{stuartIP,Calvetti2007,Kaipio2005}, where unknowns are treated as random variables, the goal of Kalman filtering is to sequentially update the posterior distribution of the unknowns conditioned on the observed data.  The classic KF estimates the model states of linear systems by updating the mean and covariance of an underlying Gaussian probability distribution through use of analytically described formulas \cite{Kalman1960}.  The EnKF accommodates use of nonlinear models by incorporating ensemble statistics into the classic KF, using an ensemble of discrete realizations from the underlying probability distribution to calculate the ensemble mean and covariance \cite{evensen,burgers}.  The augmented EnKF incorporates the simultaneous estimation of constant parameters via cross-correlation information with the model states \cite{evensen2009,arnold2014}.  The augmented EnKF can be further modified to accommodate time-varying parameter estimation through use of parameter tracking \cite{Voss2004,paramdrift,KL}. 

Each algorithm begins with a prior distribution on the system unknowns (which may include unknown model states and/or parameters) and proceeds in a predictor-corrector-type process to sequentially update the joint probability distribution of the unknowns conditioned on the available data.  Letting $\pi(x_j,\theta_j \mid y_j)$ denote the joint distribution of states $x_j$ and parameters $\theta_j$ conditioned on the data $y_j$ at time $j$, the forward model describing the state evolution, as in \eqref{eq:forward_model}, first propagates the state prediction forward to time $j+1$, forming a prediction density $\pi(x_{j+1},\theta_j \mid y_j)$.  The algorithm then corrects the prediction by using the observation model \eqref{eq:obs_model} to compare the model predictions with the available data, updating the model states and parameters to form the joint posterior distribution $\pi(x_{j+1},\theta_{j+1} \mid y_{j+1})$.  This process is repeated for each $j$ over the time span of available data.  Note that if the model parameters are known, the posterior distribution of interest simply reduces to $\pi(x_j \mid y_j)$ for state estimation.

%%%%%%%%%%%%
\subsection{Classic Kalman Filter for State Estimation}
The classic KF for state estimation begins with a Gaussian prior distribution
\begin{equation}
\pi(x_{0}) \sim\ N(\bar{x}_{0}, \mathsf{\Gamma}_{0})
\end{equation}
with mean $\bar{x}_{0} = \bar{x}_{0\mid0}\in\R^d$ and covariance matrix $\mathsf{\Gamma}_{0} = \mathsf{\Gamma}_{0\mid 0}\in\R^{d\times d}$.  The state evolution and observation equations forming the state-space model are both assumed to be linear discrete-time Markov models:
\begin{eqnarray}
X_{j+1} &=& \mathsf{F} X_{j} + V_{j+1}, \quad V_{j+1} \sim \mathcal{N}(0,\mathsf{C})  \label{eq:Classic_Forward} \\
\noalign{\vspace{.2cm}}
Y_{j+1} &=& \mathsf{G} X_{j+1} + W_{j+1}, \quad W_{j+1} \sim \mathcal{N}(0,\mathsf{D}) \label{eq:Classic_Obs}
\end{eqnarray}
for $j = 0, 1, \dots, T-1$, where the random variables $X_{j+1}$ and $Y_{j+1}$ denote the model states and observations, respectively.  The operators $\mathsf{F}$ and $\mathsf{G}$ are matrices (assumed here to be constant over time) and the noise processes $V_j$ and $W_j$ are Gaussian random variables with respective covariance matrices $\mathsf{C}$ and $\mathsf{D}$ (also assumed here to be time-invariant).

During the prediction step at time $j$, the state prediction mean $\bar{x}_{j\mid j}$, and covariance matrix $\mathsf{\Gamma}_{j\mid j}$ of the prior distribution are propagated forward in time through the use of the state evolution equation \eqref{eq:Classic_Forward} via the analytically-derived formulas
\begin{eqnarray}
\bar{x}_{j+1|j} &=& \mathsf{F} \bar{x}_{j|j} \\
\noalign{\vspace{.2cm}}
\mathsf{\Gamma}_{j+1|j} &=& \mathsf{F} \mathsf{\Gamma}_{j|j} \mathsf{F}^\mathsf{T} + \mathsf{C}
\end{eqnarray}
where $\bar{x}_{j+1|j}$ and $\mathsf{\Gamma}_{j+1|j}$ are the predicted mean and covariance of the underlying Gaussian distribution at time $j+1$ without yet taking into account the data at this point.  

In the analysis step, the predicted mean and covariance estimates are corrected using the data and the observation model \eqref{eq:Classic_Obs} via the updating formulas
\begin{eqnarray}
\bar{x}_{j+1|j+1} &=& \bar{x}_{j+1|j} + \mathsf{K}_{j+1}(y_{j+1} - \mathsf{G} \bar{x}_{j+1|j}) \\
\noalign{\vspace{.2cm}}
\mathsf{\Gamma}_{j+1|j+1} &=& (\mathsf{I} - \mathsf{K}_{j+1} \mathsf{G}) \mathsf{\Gamma}_{j+1|j}
\end{eqnarray}
where $\mathsf{I}$ is the $d\times d$ identity matrix and $\mathsf{K}_{j+1}$ is the Kalman gain matrix, defined as
\begin{equation}
\mathsf{K}_{j+1} = \mathsf{\Gamma}_{j+1|j} \mathsf{G}^\mathsf{T}(\mathsf{G}\mathsf{\Gamma}_{j+1|j} \mathsf{G}^\mathsf{T} + \mathsf{D})^{-1}.
\end{equation}
After completion of the analysis step, we are left with the Gaussian posterior distribution at time $j+1$.  The filter continues through this process, letting $j = j+1$ for $j<T$, where $T$ is the time of the last observation.

%%%%%%%%%%
\subsection{Ensemble Kalman Filter for State Estimation}  \label{sec:EnKF_state}

While the classic KF provides the optimal solution under assumptions of linearity and Gaussian distributions, these assumptions limit the algorithm's applicability to nonlinear models.  The EnKF allows for nonlinear models by incorporating ensemble statistics into the updating equations.  The state-space model in this case is given by
\begin{eqnarray}
X_{j+1} &=& F(X_{j}) + V_{j+1}, \quad V_{j+1} \sim \mathcal{N}(0,\mathsf{C})  \label{eq:Nonlin_Forward} \\
\noalign{\vspace{.2cm}}
Y_{j+1} &=& G(X_{j+1}) + W_{j+1}, \quad W_{j+1} \sim \mathcal{N}(0,\mathsf{D}) \label{eq:Nonlin_Obs}
\end{eqnarray}
where $F$ and $G$ are nonlinear operators.  In this work, $F$ is the solution to the ODE system \eqref{eq:forward_model} and $G$ is the same observation function $g$ defined in \eqref{eq:obs_model}, here with the dependence on parameters suppressed.  The EnKF maintains a similar two-step procedure of predicting and correcting, but the probability distributions are represented in terms of discrete samples, and each sample point (or ensemble member) is propagated independently.  Ensemble statistics are employed to calculate the mean and covariance of the sample at each step.

Let $S_{j\mid j}$ represent the discrete sample from the underlying probability distribution at time $j$, such that
\begin{equation}
S_{j\mid j} = \{ x_{j\mid j}^{1}, x_{j\mid j}^{2}, \dots, x_{j\mid j}^{N} \}
\end{equation}
where $N$ is the ensemble size and $x_{j\mid j}^{n}\in\R^d$ for each $n = 1,\dots,N$.  In the prediction step, each individual ensemble member is updated via the state evolution equation \eqref{eq:Nonlin_Forward} by
\begin{equation} \label{eq:enkf_pred}
x^{n}_{j+1|j} = F(x^{n}_{j|j}) + v^{n}_{j+1}, \quad n= 1,\dots,N 
\end{equation}
where each $v^{n}_{j+1}$ is a realization of the random variable $V_{j+1} \sim \mathcal{N}(0,\mathsf{C})$.  The prediction mean $\bar{x}_{j+1\mid j}$ and covariance $\mathsf{\Gamma}_{j+1\mid j}$ are then computed using the following ensemble statistics formulas:
\begin{eqnarray}
\bar{x}_{j+1\mid j} &=& \frac{1}{N} \ds\sum_{n=1}^{N} x_{j+1\mid j}^{n} \ \in \R^d \\
\noalign{\vspace{.2cm}}
\mathsf{\Gamma}_{j+1\mid j} &=& \frac{1}{N-1} \ds\sum_{n=1}^{N}  (x_{j+1\mid j}^{n} - \bar{x}_{j+1\mid j}) (x_{j+1\mid j}^{n} - \bar{x}_{j+1\mid j})^\mathsf{T} \ \in \R^{d\times d}.
\end{eqnarray}

The analysis step is performed by correcting each individual ensemble member via the equation
\begin{equation} \label{eq:enkf_analysis}
x^{n}_{j+1|j+1} =x^{n}_{j+1|j} + \mathsf{K}_{j+1}(y^{n}_{j+1} - \hat{y}^{n}_{j+1}), \quad n= 1,\dots,N 
\end{equation}
where 
\begin{equation}\label{eq:artificial_obs}
y^{n}_{j+1} = y_{j+1} + w_{j+1}^n, \quad w_{j+1}\sim\mathcal{N}(0,\mathsf{D})
\end{equation}
is an artificial observation ensemble of size $N$ generated around the data point $y_{j+1} \in \R^m$, 
\begin{equation}\label{eq:enkf_obspred}
\hat{y}^n_{j+1} = G(x^{n}_{j+1|j})  
\end{equation}
is the model prediction of the observation, and $\mathsf{K}_{j+1}$ is the Kalman gain.  To accommodate nonlinear observation models, the Kalman gain can be defined using cross-correlation information via the formula
\begin{equation}\label{eq:enkf_kg}
\mathsf{K}_{j+1} = \mathsf{\Phi}^{x\hat{y}}_{j+1} (\mathsf{\Phi}^{\hat{y}\hat{y}}_{j+1} + \mathsf{D})^{-1}
\end{equation}
where $\mathsf{\Phi}^{x\hat{y}}_{j+1}$ is the cross covariance of the state and observation predictions, $\mathsf{\Phi}^{\hat{y}\hat{y}}_{j+1}$ is the forecast error of the observation prediction ensemble, and $\mathsf{D}$ is the observation noise covariance \cite{Morad2005}.  This results in the posterior ensemble
\begin{equation}
S_{j+1\mid j+1}=\{ x_{j+1\mid j+1}^{1}, x_{j+1\mid j+1}^{2}, \dots, x_{j+1\mid j+1}^{N} \}
\end{equation}
and the corresponding posterior ensemble mean $\bar{x}_{j+1\mid j+1}$ and covariance $\mathsf{\Gamma}_{j+1\mid j+1}$ computed via ensemble statistics.  As in the classic KF, this process continues while $j <T$.

%%%%%%%%
\subsection{Augmented EnKF for State and Constant Parameter Estimation} \label{sec:AEnKF_param}

While the EnKF allows for more flexibility in accommodating nonlinear models for both the states and observations, it can be further modified to simultaneously estimate both model states and constant parameters through use of augmented vectors.  The state-space model becomes
\begin{eqnarray}
X_{j+1} &=& F(X_{j},\theta) + V_{j+1}, \quad V_{j+1} \sim \mathcal{N}(0,\mathsf{C})  \label{eq:NonlinParam_Forward} \\
\noalign{\vspace{.2cm}}
Y_{j+1} &=& G(X_{j+1},\theta) + W_{j+1}, \quad W_{j+1} \sim \mathcal{N}(0,\mathsf{D}) \label{eq:NonlinParam_Obs}
\end{eqnarray}
where $\theta\in\R^p$ denotes the unknown system parameters, and the goal is to estimate a joint probability distribution representing the model states and parameters conditioned on the available data.  To this end, the discrete sample from the probability distribution at time $j$ becomes
\begin{equation}
S_{j\mid j} = \{ (x_{j\mid j}^{1}, \theta_{j\mid j}^1), (x_{j\mid j}^{2}, \theta_{j\mid j}^2), \dots, (x_{j\mid j}^{N}, \theta_{j\mid j}^N) \}
\end{equation}
where each ensemble member is now a joint state-parameter sample.  The prediction step of the filter remains largely the same, propagating the model states forward via the state evolution equation \eqref{eq:NonlinParam_Forward}, such that 
\begin{equation}\label{eq:enkf_pred_aug}
x^{n}_{j+1|j} = F(x^{n}_{j|j},\theta_{j\mid j}^n) + v^{n}_{j+1}, \quad n= 1,\dots,N 
\end{equation}
while the parameter estimates remain unaltered, with
\begin{equation}
\theta_{j+1\mid j}^n = \theta_{j\mid j}^n
\end{equation}
for each $n = 1,\dots, N$.

After completion of the prediction step, the state predictions and parameters are augmented to form the joint sample vectors
\begin{equation}
z_{j+1\mid j}^{n} = \begin{bmatrix} x^{n}_{j+1\mid j} \\ \theta_{j+1\mid j}^{n}\end{bmatrix} \ \in\R^{d+p}, \quad n = 1,\dots, N.
\end{equation}
The prediction ensemble mean and covariance are then computed using ensemble statistics via the joint sample vectors, thereby obtaining correlation information between the model states and parameters that is used in the analysis step
\begin{equation} \label{eq:enkf_analysis_aug}
z^{n}_{j+1|j+1} =z^{n}_{j+1|j} + \mathsf{K}_{j+1}(y^{n}_{j+1} - \hat{y}^n_{j+1}), \quad n= 1,\dots,N 
\end{equation}
to simultaneously update the state and parameter estimates, where the cross-correlation information is encoded in the Kalman gain.

%%%%%%%%
\subsection{Augmented EnKF for Time-Varying Parameter Tracking} \label{sec:AEnKF_TVparam}

While the augmented EnKF is commonly employed for estimating constant parameters, it can be modified in the prediction step to track time-varying parameters if the parameters change more slowly than the system states.  More specifically, during the prediction step of the filter, time-varying parameters are updated using a random walk of the form
\begin{equation}\label{Eq:ParamDrift}
\theta_{j+1|j}^n  = \theta_{j|j}^n + \xi_{j+1}^n, \quad n = 1, \dots, N
\end{equation}
where $\xi_{j+1}^n \sim \mathcal{N}(0,\mathsf{E})$ defines the parameter drift.  Parameter tracking allows the filter to estimate time-varying functions without a priori assumptions on functional form.  The drift covariance matrix $\mathsf{E}$, which is set in advance of running the filter, plays a vital role in the algorithm's ability to successfully track the time-varying parameter of interest without diverging \cite{KL}.

%%%%%%%%%%%%%%%%%%%%%%%%%%%%%%%%%%%%%%%%%%%%%%%%%%
\section{Observation Functions for Epidemic Modeling} \label{ch:SIR}

In this section we describe both the forward model and the observation functions used in the EnKF framework for our epidemic application.  As seen in Section~\ref{ch:kf}, the observation function plays a significant role in connecting the filter estimates in the prediction step back to the data in the analysis step.  For our analysis in this work, we model the forward dynamics of the system using the Susceptible-Infectious-Recovered (SIR) model detailed below, along with four different observation functions of increasing complexity used in epidemic modeling.

%%%%%%%%
\subsection{Forward Model: SIR}

\begin{figure}[t!]
\centering
\includegraphics[width=.4\linewidth]{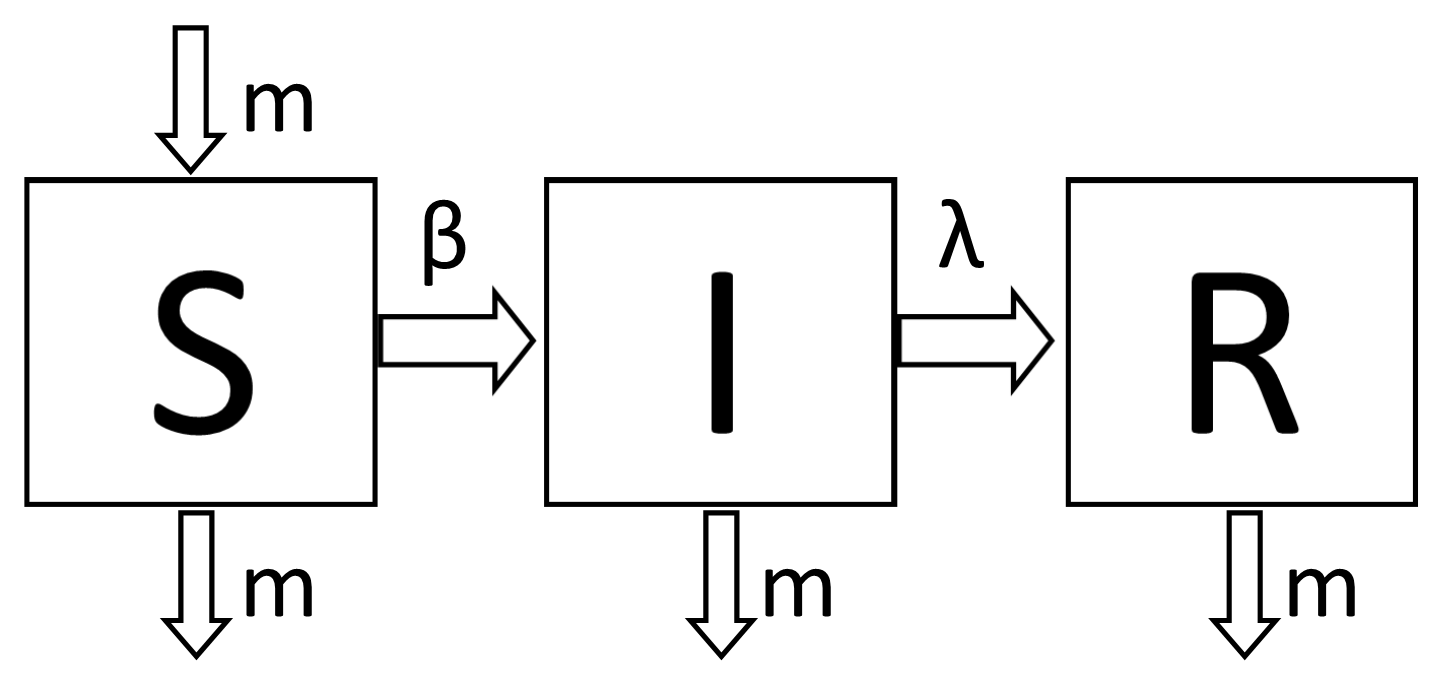}
\caption{A diagram of the Susceptible-Infectious-Recovered compartment model, where $m$ is a constant birth and death rate, $\beta(t)$ is the time-varying transmission parameter, and $\lambda$ is the constant recovery rate.  The total population size $N_\text{pop}$ is assumed to be constant, such that $N_\text{pop} = S(t) + I(t) + R(t)$. }
\label{fig:SIR}
\end{figure}

As noted in the introduction, the SIR model is a classic model in epidemiology \cite{KermackSIR,DC,yellowbook,greenbook}.  The standard form of the model comprises three compartments describing three different states of a population at a given time $t$: the susceptible population, $S(t)$, are healthy with the chance of contracting the disease; the infectious population, $I(t)$, are able to transmit the disease to others; and the recovered population, $R(t)$, are those who have recovered from and have become immune to the disease \cite{KermackSIR}.  Figure~\ref{fig:SIR} gives a schematic representation of the SIR model used in this work.  Assuming a constant population size $N_\text{pop}$, the recovered population can be written as a function of the susceptible and infectious populations, so that $R(t) = N_\text{pop} - S(t) - I(t)$.  

The corresponding system of ODEs describing the rate of change of the susceptible and infectious populations is given by
\begin{eqnarray}
\frac{dS(t)}{dt} &=& mN_\text{pop} - \frac{\beta(t) I(t)S(t)}{N_\text{pop}} -mS(t) \label{eq:SIR1}\\
\noalign{\vspace{.2cm}}
\frac{dI(t)}{dt} &=& \frac{\beta(t) I(t)S(t)}{N_\text{pop}} - \lambda I(t) -mI(t) \label{eq:SIR2}
\end{eqnarray}
where $m$ is a constant birth and death rate, $\beta(t)$ is the transmission parameter at time $t$, and $\lambda$ is the constant recovery rate.  The solution to the system \eqref{eq:SIR1}--\eqref{eq:SIR2} defines the forward propagation in the prediction step of the filter; i.e., the function $F$ in \eqref{eq:enkf_pred} for state estimation and \eqref{eq:enkf_pred_aug} for combined state and parameter estimation, where $x = (S, I) \in \R^2$ and $\theta$ denotes the unknown model parameters.

%%%FIGURE
%%%

%%%%%%%%
\subsection{Observation Functions: Prevalence, Incidence, and Under-Reporting}

Here we describe the four epidemiologically-inspired observation functions considered in this work, which range from assuming direct measurements of the infectious population (prevalence data) to assuming an under-reported number of newly infected individuals over a specified time period (under-reported incidence data).  Recall that the observation function $g(x,\theta)$ in \eqref{eq:obs_model} is responsible for relating the observed data back to the model predictions in the analysis step of the Kalman filtering algorithms.  More specifically, this function is used to compute the observation predictions in \eqref{eq:enkf_obspred}, which are then compared with the available data in either \eqref{eq:enkf_analysis} for the EnKF or \eqref{eq:enkf_analysis_aug} for the augmented EnKF.  In a given application, the observation function aims to model how the set of observed data corresponds to the forward model variables.  

To illustrate the effects of observation model selection on the results of the filtering algorithms described in Section~\ref{ch:kf}, we employ the following four observation functions of varying complexity, representing four different data collection assumptions: 
\begin{enumerate}
% Case 1
\item \textbf{Prevalence Data:}
The observed data at time $t_j$ is assumed to be a direct measurement of the infectious population at time $t_j$:
\begin{equation} \label{eq:case1}
g(x_{j}, \theta) = I(t_j) 
\end{equation}

% Case 2
\item \textbf{Under-Reported Prevalence Data:}
The observed data at time $t_j$ is assumed to be an under-reported measure of the infectious population at time $t_j$:
\begin{equation} \label{eq:case2}
g(x_{j}, \theta) = \rho I(t_j)
\end{equation}
Here the constant parameter $\rho$ is the reporting probability, which denotes the percentage of the population actively reporting cases.  This function accounts for the fact that not every infectious person will report their illness, and thus data sets may not be complete. 

% Case 3
\item \textbf{Incidence Data:}
The observed data at time $t_j$ is assumed to be a measurement of the total number of cases accumulated from time $t_{j-1}$ to time $t_j$:
\begin{equation} \label{eq:case3}
g(x_{j}, \theta) = \int_{t_{j-1}}^{t_{j}} \frac{\beta(t)I(t)S(t)}{N_\text{pop}}dt
\end{equation}

% Case 4
\item \textbf{Under-Reported Incidence Data:}
The observed data at time $t_j$ is assumed to be an under-reported measure of the total number of cases accumulated from time $t_{j-1}$ to time $t_j$:
\begin{equation} \label{eq:case4}
g(x_{j}, \theta) = \rho \int_{t_{j-1}}^{t_{j}} \frac{\beta(t)I(t)S(t)}{N_\text{pop}}dt
\end{equation}
where, as in \eqref{eq:case2}, $\rho$ is the reporting probability.
\end{enumerate}
In the following sections, we refer to the four observation functions in \eqref{eq:case1}--\eqref{eq:case4} as Cases 1--4, respectively.  Note that the functions in Cases 1 and 2 are linear with respect to the model states, while the functions in Cases 3 and 4 are nonlinear.

%%%%%%%%
\subsection{Overview of Simulation Studies}

In the numerical experiments that follow, we show how inadvertent use of a suboptimal observation function (e.g., using a prevalence-based model for incidence data or neglecting to account for under-reporting) leads to substandard filtering performance.  More specifically, we simulate under-reported monthly incidence data using the Case 4 observation function, and then we test the effects of using each of the observations functions in Cases 1--4 together with the SIR model \eqref{eq:SIR1}--\eqref{eq:SIR2} in three filtering scenarios of increasing complexity: the EnKF for state estimation with known parameters; the augmented EnKF for combined state and constant parameter estimation; and the augmented EnKF with parameter tracking for combined state and time-varying parameter estimation.

%%%%%%%%%%%%%%%%%%%%%%%%%%%%%%%%%%%%%%%%%%%%%%%%%%%%%
\section{Numerical Results}\label{ch:results}

In this section, we perform simulation studies to demonstrate the effects of observation function selection on the resulting EnKF estimates by incorporating the four observation functions from Section \ref{ch:SIR} into the EnKF framework to assimilate synthetic data generated using the under-reported incidence observation function (i.e., Case 4).  We analyze the results in three different filtering scenarios, including state estimation with known parameters and combined state and parameter estimation with unknown constant and time-varying parameters.  We further illustrate how the observation noise covariance matrix $\mathsf{D}$ in the filter affects the results.  All numerical experiments were performed using MATLAB\textsuperscript{\textregistered} programming language (The MathWorks, Inc., Natick, MA).  The filtering algorithms were hand coded in MATLAB using \texttt{ode15s} to numerically solve the ODE system \eqref{eq:SIR1}--\eqref{eq:SIR2} in the prediction step.

%%%%%%%%
\subsection{Filter Initialization, Data Generation, and Outline of Experiments}\label{subsec:data_gen}

For each numerical experiment that follows, we used $N=100$ ensemble members to represent the underlying probability distributions.  While not shown, we performed additional numerical tests to ensure that the results obtained using 100 ensemble members were consistent with those using larger ensemble sizes.  Initial ensembles for the state and parameter values were drawn from uniform prior distributions containing but not centered at the true initial values and parameter values used in generating the simulated data.  For the numerical experiments in Sections~\ref{subsec:state_est} and \ref{subsec:param_est}, we assign the model noise covariance matrix to be $\mathsf{C} = \sigma_{C}^2 \mathsf{I}_2$, where $\mathsf{I}_2$ is the $2\times2$ identity matrix, with $\sigma_{C} = 0.2$ and the observation noise covariance matrix to be $\mathsf{D} = \sigma_{D}^2$ with $\sigma_{D} = 1$.  In Section~\ref{subsec:incD} we increase the value of $\sigma_{D}$ to include additional error accounting for potential use of suboptimal observation functions.

Simulated data was generated using the under-reported incidence observation function in \eqref{eq:case4} (i.e., Case 4) and the parameter values in Table~\ref{Tab:Params}, which are comparable to parameter values used in modeling the spread of measles; a similar example was considered in \cite{arnold2018}.  The time-varying transmission parameter $\beta(t)$ was modeled as
\begin{equation} \label{eq:transmission}
\beta(t) = b _{0}\big(1 + b_{1}\cos(2\pi t)\big)
\end{equation}
with constant average transmission $b_0$ and amplitude $b_1$.  Assuming that 95\% of the population was initially susceptible and 2\% initially infected, we solved the SIR model \eqref{eq:SIR1}--\eqref{eq:SIR2} numerically using \texttt{ode45} to first reach a steady state, then restarted the model simulation for data collection.  Data consist of monthly observations of an under-reported number of newly infected individuals over a time span of 10 years, with 70\% of cases reported each month, corrupted by a small amount of Gaussian noise with zero mean and standard deviation 0.1.  We note that this noise model assumes constant error variance regardless of the magnitude of cases; for discussion of an alternative approach, see Appendix~\ref{Appendix:Noise}.  Figure~\ref{fig:synth_data} plots the observed data, along with the true solution curves for $S(t)$ and $I(t)$.

%%%%TABLE
\begin{table}[t!]  
\begin{center}
\begin{tabular}{|p{2cm}||p{4cm}|p{2cm}|p{2cm}|}
 \hline
 \multicolumn{4}{|c|}{{Parameters Used in Synthetic Data Generation}} \\
 \hline
Notation & Meaning & Value & Units\\
 \hline
$N_\text{pop}$ & Population Size & 90,000 & individuals \\
$b_{0} $ & Average Transmission & 1800 & 1/years \\
$b_{1} $ & Amplitude & 0.08 & 1/years \\
$\lambda$ & Recovery Rate & 100 & 1/years \\
$m$ & Birth/Date Rate & 0.02 & 1/years \\
$\rho$ & Reporting Probability & 0.70 & -- \\
 \hline
\end{tabular}
\end{center}
\caption{True parameter values used in generating the synthetic data in Figure~\ref{fig:synth_data}, mimicking the spread of measles.  The transmission parameter $\beta(t)$ is modeled as in \eqref{eq:transmission} with average transmission $b_0$ and amplitude $b_1$.  
\label{Tab:Params}}
\end{table} 
%%%%%

Given the observed data described above, we perform the following numerical experiments: In Section~\ref{subsec:state_est} we test how using the four different observation functions affects the EnKF for state estimation in accurately estimating the time series of both the susceptible and infectious populations.  In Section~\ref{subsec:param_est} we study the effects when using the augmented EnKF to estimate the model states along with the constant parameters $b_0$ and $b_1$ relating to the functional form of the transmission parameter $\beta(t)$ in \eqref{eq:transmission}.  We then analyze results when using the augmented EnKF with parameter tracking to estimate the full time series of the transmission parameter $\beta(t)$ without assuming a functional form.  Finally, in Section~\ref{subsec:incD} we illustrate how the observation noise covariance matrix $\mathsf{D}$ in the filter can be utilized in certain cases to help account for uncertainty in the observation function selection.  We note while the data is simulated (and thus the underlying true observation function is known), we do not assume to have knowledge of the true observation function in addressing these inverse problems.  

We present the results for each experiment in figures plotting the time series estimates of $S(t)$, $I(t)$, the monthly number of cases (when applicable), and any estimated parameters (as relevant).  Along with the plots, we also compute the mean squared error (MSE) of the model states in each of the trials by calculating the average of the squared differences between the true and estimated susceptible and infectious populations at each time step:
\begin{eqnarray}
MSE_{S} &=& \frac{1}{M} \sum_{i = 1}^{M} (S^\text{true}_{i} - S^\text{est}_{i})^{2} \\
\noalign{\vspace{.2cm}}
MSE_{I} &=& \frac{1}{M} \sum_{i = 1}^{M} (I^\text{true}_{i} - I^\text{est}_{i})^{2}
\end{eqnarray}
where $S^\text{true}$ and $I^\text{true}$ are the true model states in Figure~\ref{fig:synth_data}, and $S^\text{est}$ and $I^\text{est}$ are the EnKF mean estimates of the states, respectively.  Here $M$ is the total number of time points of comparison, which in our simulations equals the number of observations.  An additional check for filter consistency is discussed in Appendix~\ref{Appendix:Consistency}.

%%%FIGURE
\begin{figure}[t!]
\centering
\includegraphics[width = \textwidth]{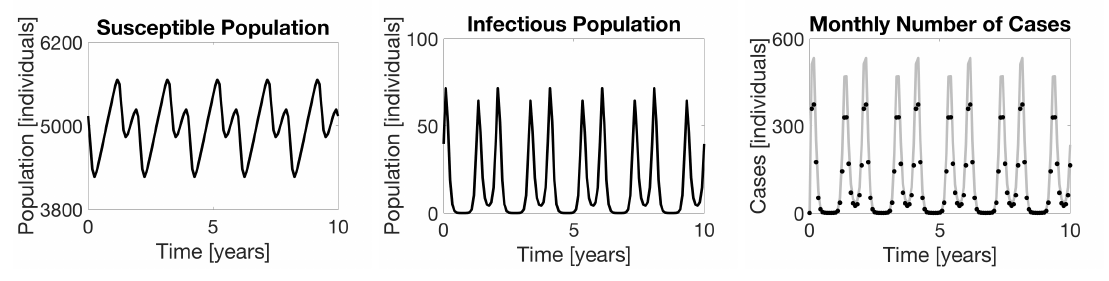}
\caption{ Synthetic data generated from the SIR model \eqref{eq:SIR1}--\eqref{eq:SIR2} with parameters given in Table~\ref{Tab:Params}.  The plots show the true susceptible (left) and infectious (middle) populations, along the with true number of monthly cases (right, gray) and the observed data (right, black markers) generated using the Case 4 observation function \eqref{eq:case4} at 70\% reporting and corrupted with Gaussian noise. }
\label{fig:synth_data}
\end{figure}
%%%%%

%%%%%%%%%%%%%%%%%%%%
\subsection{State Estimation with Known Parameters}\label{subsec:state_est}

\begin{figure}[p]
\centering
\includegraphics[width = \textwidth]{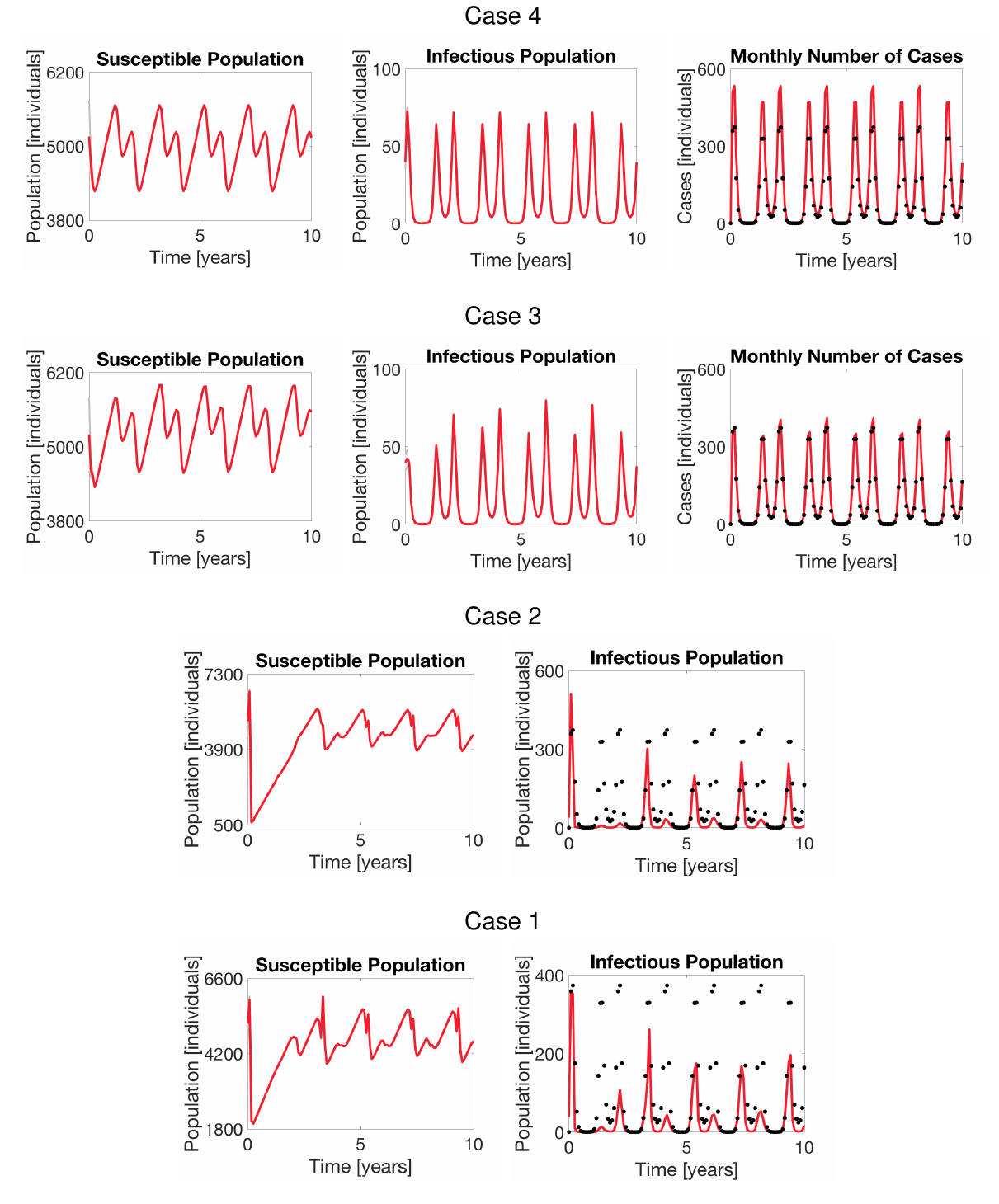}
\caption{Results of the EnKF for state estimation with known parameters when employing (from top to bottom): the Case 4 observation function \eqref{eq:case4}; the Case 3 observation function \eqref{eq:case3}; the Case 2 observation function \eqref{eq:case2}; and the Case 1 observation function \eqref{eq:case1}.  In each plot, the EnKF mean is shown in solid red, the $\pm2$ standard deviation curves around the mean are filled in gray, and the observed data are plotted in black markers.   }
\label{fig:og}
\end{figure}

Assuming known values for the parameters, as given in Table~\ref{Tab:Params}, we employ the EnKF for state estimation (outlined in Section~\ref{sec:EnKF_state}) to estimate $S(t)$, $I(t)$, and the monthly number of cases when applicable.  Note that when using the observation functions in Cases 3 and 4, the monthly number of cases is computed as a separate function; in Cases 1 and 2, the data is interpreted instead as a direct percentage of the estimated $I(t)$.  Figure \ref{fig:og} shows the results when using the observation functions in each of the four cases, and Table~\ref{Tab:MSE_OG} lists the corresponding MSE values.

%%%% FIGURE
%%%%

Since the observation function in Case 4 is the true function used in generating the synthetic data, it is not surprising that the filter performs well when using this function.  As seen in Figure~\ref{fig:og}, in this case the EnKF is able to accurately estimate all components of the model, with the EnKF mean estimates nearly identical to the underlying true model states.  Although the estimated $\pm2$ standard deviation curves are plotted, the standard deviations around the mean are so small that they are not easily visible in the plots, suggesting a high level of confidence in the mean estimates.  This test validates that the filter is able to well estimate the corresponding model states when using an observation function that well interprets the available data. 

By assuming full reporting in the Case 3 observation function, some detrimental effects on the corresponding filter estimates already become clear.  In particular, since the Case 3 function does not account for under-reporting, the filter fits the monthly number of cases more directly to the observed data points.  While the estimated $S(t)$ and $I(t)$ curves follow the same general shape as in the original model, the magnitudes of the estimates are generally higher for $S(t)$ with alternating higher peaks for $I(t)$.  These results are further evident in the MSE values recorded in Table \ref{Tab:MSE_OG}. 

When using the observation functions in Cases 1 and 2, we see more significant effects due to the interpretations of the data: these functions interpret the observed data as either direct observations of the infectious population (Case 1) or a proportion of the population (Case 2) at a given time, essentially misinterpreting incidence data as prevalence.  The results in Figure~\ref{fig:og} show that when using the observation function in Case 2, neither the shape nor magnitude of the EnKF estimates resemble the true model states.  Similar results follow when using the observation function in Case 1, with discrepancies in both the shape and magnitude of the underlying true curves.  For each of these cases, the corresponding MSE values are noticeably high.  In particular, while the MSE values for $S(t)$ are of similar magnitude as in Case 3, the errors are much larger for $I(t)$.

%%% TABLE
\begin{table}[t!]
\begin{center}
\begin{tabular}{ |p{1cm}||p{3cm}|p{3cm}|}
 \hline
 \multicolumn{3}{|c|}{ MSE for State Estimation} \\
 \hline
Case & $MSE_S$ & $MSE_I$ \\
 \hline
1  & $7.89 \times 10^{5}$ & $2.79 \times 10^{3} $\\
2 &  $1.86 \times 10^{6}$ & $4.07 \times 10^{3}$\\
3 & $1.01 \times 10^{5}$ & $16.28$\\
4 & 22.30 & 0.01 \\
 \hline
\end{tabular}
\end{center}
\caption{Mean squared error of the model states $S(t)$ and $I(t)$ when using the EnKF for state estimation with known parameters and the observation functions in Cases 1--4, respectively.  MSE values are reported to two decimal places.  }
\label{Tab:MSE_OG}
\end{table}
%%%%

%%%%%%%%%%%%%%%%%%%%%%
\subsection{Combined State and Parameter Estimation}\label{subsec:param_est}

In this section we consider the augmented EnKF for combined state and parameter estimation.  More specifically, we analyze the effects of observation function selection when estimating both constant and time-varying parameters relating to the disease transmission $\beta(t)$, modeled as in \eqref{eq:transmission}.  For constant parameter estimation, we assume the form of $\beta(t)$ in \eqref{eq:transmission} and aim to estimate the constants $b_0$ and $b_1$, representing the average transmission and amplitude of variation, respectively.  For time-varying parameter estimation, we do not assume a known form for $\beta(t)$ and instead use parameter tracking to approximate the time series.  In both scenarios, we assume that the remaining model parameters are known and set to the values in Table~\ref{Tab:Params}.

%%%%%%%%
\subsubsection{Constant Parameter Estimation}

\begin{figure}[h!]
\centering
\includegraphics[width = \textwidth]{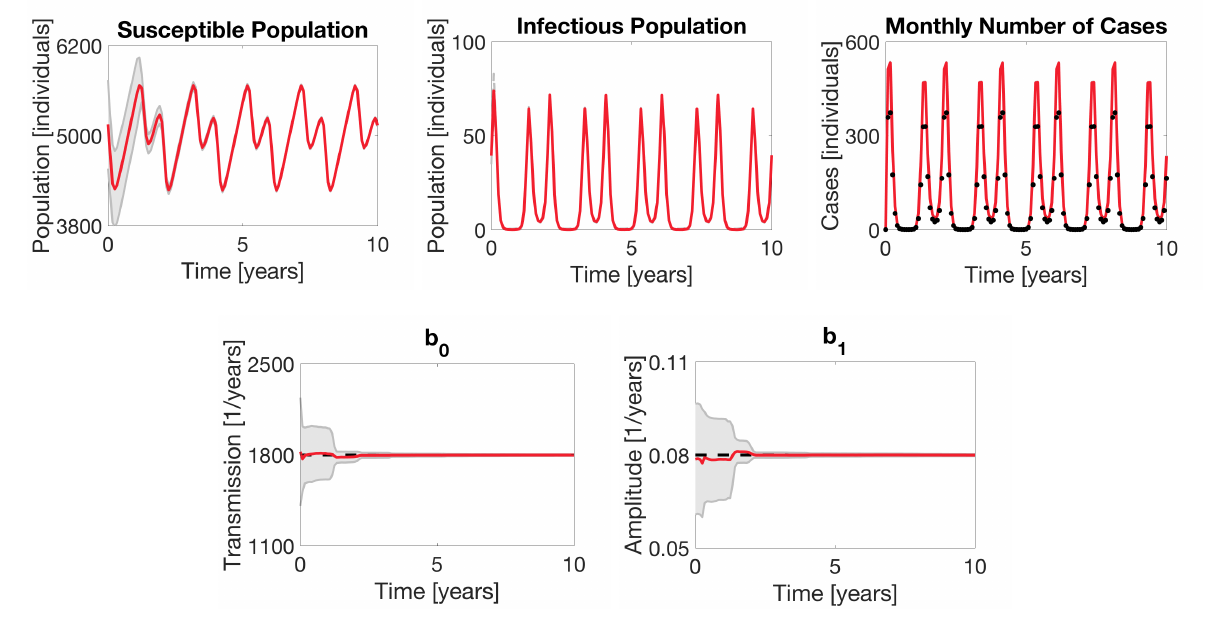}
\caption{
Results of the augmented EnKF for combined state and constant parameter estimation when employing the Case 4 observation function \eqref{eq:case4}.  In each plot, the EnKF mean is shown in solid red, the $\pm2$ standard deviation curves around the mean are filled in gray, and the observed data are plotted in black markers.  The true parameter values for $b_0$ and $b_1$ are plotted in dashed black.  Around year 2, the filter converges closely to the true values of $b_{0}$ and $b_{1}$. 
}
\label{fig:case4_aug}
\end{figure}

\begin{figure}[h!]
\centering
\includegraphics[width = \textwidth]{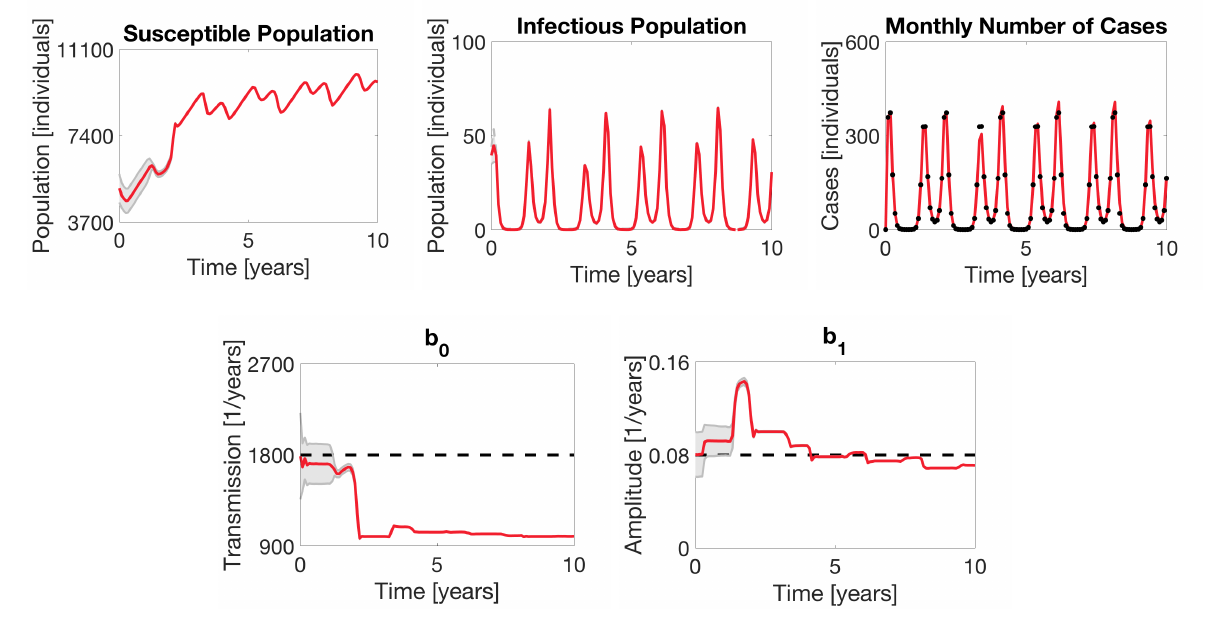}
\caption{
Results of the augmented EnKF for combined state and constant parameter estimation when employing the Case 3 observation function \eqref{eq:case3}.  In each plot, the EnKF mean is shown in solid red, the $\pm2$ standard deviation curves around the mean are filled in gray, and the observed data are plotted in black markers.  The true parameter values for $b_0$ and $b_1$ are plotted in dashed black.  Not accounting for under-reporting in the observation function causes the parameter estimation to become more difficult for the filter.  
}
\label{fig:case3_aug}
\end{figure}

Here we apply the augmented EnKF (outlined in Section~\ref{sec:AEnKF_param}) to estimate $S(t)$, $I(t)$, the monthly number of cases when applicable, and the constant parameters $b_{0}$ and $b_{1}$ from the transmission function \eqref{eq:transmission}.  Table~\ref{tbl:MSE_aug} lists the MSE values when employing the observation functions in Cases 1--4, and Figures \ref{fig:case4_aug} and \ref{fig:case3_aug} show the results for Cases 4 and 3, respectively. 

As in the previous experiment, the augmented EnKF using Case 4 is able to estimate the true model states and unknown parameters with high accuracy.  As shown in Figure~\ref{fig:case4_aug}, the initial uncertainty around the estimates shrinks around year 2, when the parameter estimates converge to their true values.  The resulting EnKF posterior mean estimate for $b_0$ converges to the true value with a relative error of $1.83\times 10^{-4}$ and for $b_1$ with a relative error of $4.49\times 10^{-4}$.

%%%TABLE
\begin{table}[t!]
\begin{center}
\begin{tabular}{ |p{1cm}||p{3cm}|p{3cm}|}
 \hline
 \multicolumn{3}{|c|}{MSE for Constant Parameter Estimation} \\
 \hline
 Case & $MSE_S$ & $MSE_I$ \\
 \hline
1  & $1.05 \times 10^{8}$ & $2.31 \times 10^{3}$  \\
2 & $1.03 \times 10^{8}$ & $3.86 \times 10^{3}$ \\
3 & $1.30 \times 10^{7} $&  $49.21$ \\
4  & $235.88$ & $0.06$ \\
 \hline
 \end{tabular}
\end{center}
\caption{
Mean squared error of the model states $S(t)$ and $I(t)$ when using the augmented EnKF for combined state and constant parameter estimation and the observation functions in Cases 1--4, respectively.  MSE values are reported to two decimal places.  The increase in complexity of the inverse problem results in an overall increase in MSE values; yet the error remains smaller when using observation functions closer in form to the true function in Case 4.
}
\label{tbl:MSE_aug}
\end{table}
%%%%

%%%% FIGURE
%%%%

%%%% FIGURE
%%%%

However, not accounting for under-reporting when using the Case 3 observation function causes the filter to have much more difficulty in estimating $b_0$ and $b_1$.  Figure~\ref{fig:case3_aug} shows a drop in magnitude for the estimate of $b_{0}$, resulting in a posterior estimate close to half of the true value.  There is more instability in the estimate for $b_{1}$, resulting in a posterior estimate not as far from the true value but not fully converged.  In this case, the relative error in the posterior estimate for $b_0$ is $4.47\times10^{-1}$ and for $b_1$ is $1.12\times10^{-1}$, respectively.  The effects of the inaccurate parameter estimates are also seen in the state estimates, most noticeably in the estimate for $S(t)$ shown in Figure~\ref{fig:case3_aug} and in the corresponding MSE listed in Table~\ref{tbl:MSE_aug}.   

Similar to the state estimation experiments, constant parameter estimation results continue to degrade when using the Case 1 and Case 2 observation functions.  While not shown here to avoid redundancy, in both of these cases the filter is unable to find the true values of $b_{0}$ and $b_{1}$ and the estimates do not converge, diverging more drastically than in the previous case.  While the MSE values in Table~\ref{tbl:MSE_aug} for $I(t)$ stay within the same magnitude for Cases 1 and 2, there is a large increase in these values as compared to Cases 3 and 4.

%%%%%%%%
\subsubsection{Time-Varying Parameter Estimation}

\begin{figure}[t!]
\centering
\includegraphics[width = \textwidth]{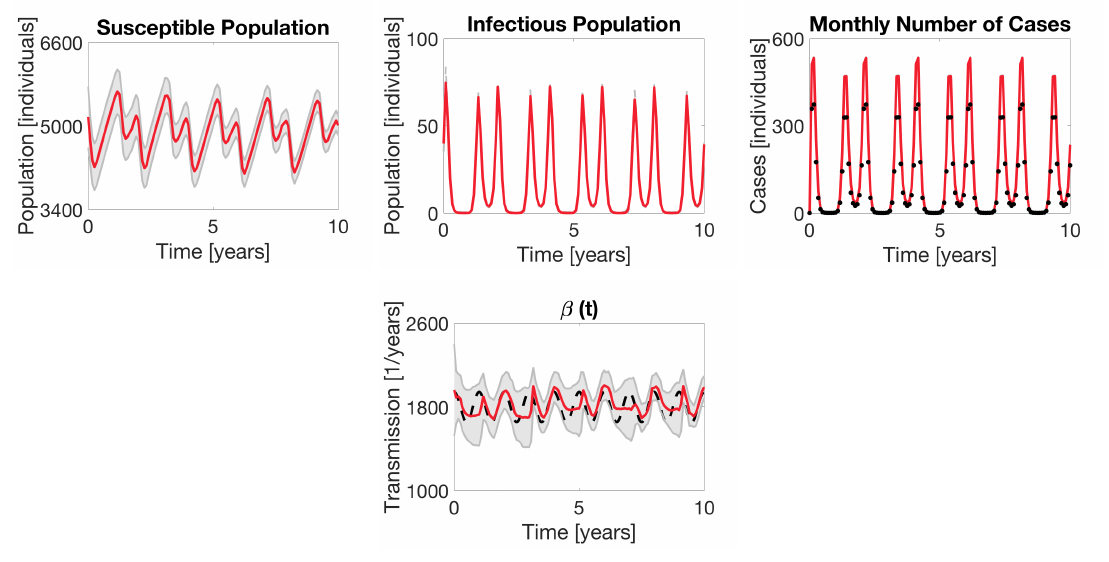}
\caption{
Results of the augmented EnKF with parameter tracking for combined state and time-varying parameter estimation when employing the Case 4 observation function \eqref{eq:case4}.  In each plot, the EnKF mean is shown in solid red, the $\pm2$ standard deviation curves around the mean are filled in gray, and the observed data are plotted in black markers.  The true functional form of the time-varying transmission parameter $\beta(t)$ is plotted in dashed black.  Estimating the time-varying transmission parameter is a challenging problem even when using the true observation function.  
}
\label{fig:case4_pd}
\end{figure}

\begin{figure}[h!]
\centering
\includegraphics[width = \textwidth]{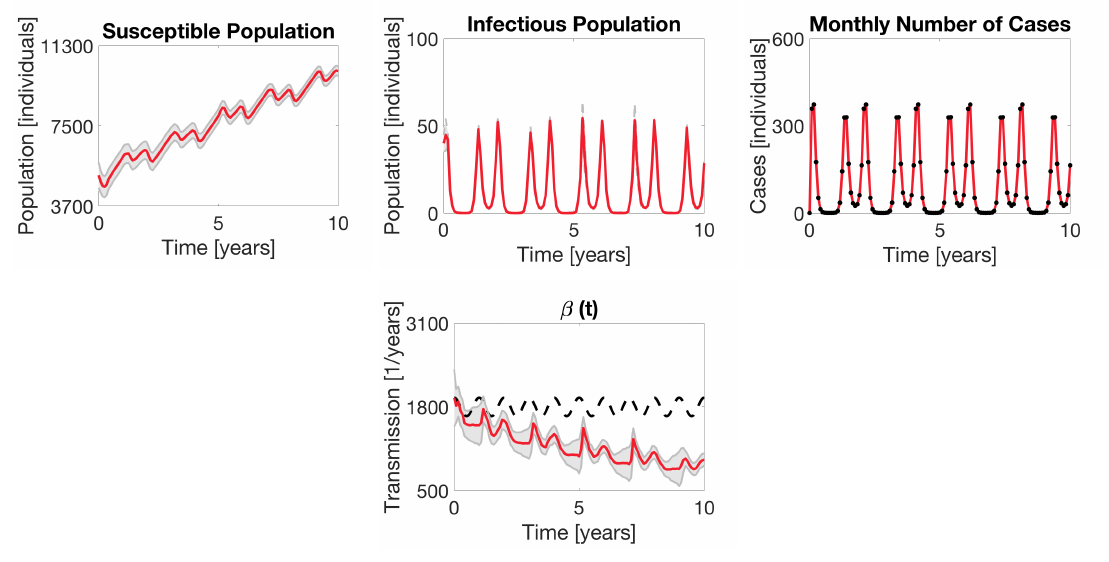}
\caption{
Results of the augmented EnKF with parameter tracking for combined state and time-varying parameter estimation when employing the Case 3 observation function \eqref{eq:case3}.  In each plot, the EnKF mean is shown in solid red, the $\pm2$ standard deviation curves around the mean are filled in gray, and the observed data are plotted in black markers.  The true functional form of the time-varying transmission parameter $\beta(t)$ is plotted in dashed black.  Not accounting for under-reporting in the observation function makes an already challenging problem more difficult. }
\label{fig:case3_pd}
\end{figure}

In this section we apply the augmented EnKF with parameter tracking (outlined in Section~\ref{sec:AEnKF_TVparam}) to estimate $S(t)$, $I(t)$, the monthly number of cases when applicable, and the time-varying transmission parameter $\beta(t)$.  We set the drift covariance matrix in \eqref{Eq:ParamDrift} to be $\mathsf{E} = \sigma_{E}^2$ with $\sigma_{E} = 45$.  While the true $\beta(t)$ used in generating the data has the form given in \eqref{eq:transmission}, in this setting we do not assume a known form of the parameter and instead use a random walk to track the parameter time series as the algorithm progresses.  Table~\ref{tbl:MSE_pd} lists the MSE values when employing the observation functions in Cases 1--4, and Figures \ref{fig:case4_pd} and \ref{fig:case3_pd} show the results for Cases 4 and 3, respectively. 

By not assuming a known form for $\beta(t)$, the time-varying parameter estimation presents the most challenging of the filtering problems considered in this work.  As shown in Figure~\ref{fig:case4_pd}, even using the Case 4 observation function (the true observation function used in generating the data) does not result in a highly accurate reconstruction of the time series of $\beta(t)$.  While the periodicity of the underlying sinusoidal function is not well captured, the parameter tracking estimate of $\beta(t)$ does remain within the same range of values and fully captures the true function within the $\pm2$ standard deviation curves around the mean.  The corresponding EnKF estimates of the model states and monthly number of cases still remain relatively accurate, with noticeably larger uncertainty bounds for $S(t)$.

%%%TABLE
\begin{table}[t!]
\begin{center}
\begin{tabular}{ |p{1cm}||p{3cm}|p{3cm}|}
 \hline
 \multicolumn{3}{|c|}{MSE for Time-Varying Parameter Estimation} \\
 \hline
 Case & $MSE_S$ & $MSE_I$ \\
 \hline
1  & $4.73 \times 10^{7}$ & $2.60 \times 10^{3}$  \\
2 & $1.0 \times 10^{8}$ & $3.85 \times 10^{3}$ \\
3 & $9.08 \times 10^{6}$ & $58.55$ \\
4  & $1.62 \times 10^{4}$ & $0.60$ \\
 \hline
\end{tabular} 
\end{center}
\caption{
Mean squared error of the model states $S(t)$ and $I(t)$ when using the augmented EnKF with parameter for combined state and time-varying parameter estimation and the observation functions in Cases 1--4, respectively.  MSE values are reported to two decimal places.  The increase in complexity of the inverse problem again causes an overall increase in MSE values; yet the errors remain smaller when using the observation functions in Cases 3 and 4.}
\label{tbl:MSE_pd}
\end{table}
%%%%

%%%% FIGURE
%%%%

%%%% FIGURE
%%%%

Similar to the constant parameter estimation experiments, using the observation function in Case 3 here degrades the filter's performance in estimating $\beta(t)$ as well as the corresponding model states.  As shown in Figure~\ref{fig:case3_pd}, while somewhat maintaining the sinusoidal shape, the estimate of the transmission function begins to drift downward outside a reasonable range of the true function values.  This accounts for a corresponding increase in the susceptible population estimate and underestimation of the infectious population and monthly number of cases.  

Consistent with the previous experiments, using the observation functions in Cases 1 and 2 causes further degradation in performance, with the filter unable to well track the underlying transmission parameter, thereby resulting in similarly poor estimates of the model states.  This is reflected in the resulting MSE values listed in Table~\ref{tbl:MSE_pd}.

%%%%%%%%%%%%%%%%%%%%%%%%%%%%%%%%%%%%%%%%%
\subsection{Observation Noise Covariance}\label{subsec:incD}

The numerical experiments conducted in the previous sections demonstrate how using different observation functions for the same observed data within the EnKF framework affects the resulting estimates for both the model states and parameters.  More specifically, the results show that using an observation function that inaccurately interprets the epidemiological data (i.e., mistaking incidence for prevalence, or neglecting under-reporting) can lead to inaccurate estimates of the system unknowns.  In this section, we illustrate how the observation noise covariance matrix $\mathsf{D}$ in the filter can be utilized in certain cases to offset some of the detrimental effects of using a suboptimal observation function.  

In the observation model \eqref{eq:obs_model}, the observed data $y_j$ are assumed to be outputs from the observation function $g$ corrupted by additive noise, where the error is normally distributed with zero mean and covariance $\mathsf{D}$.  The noise covariance is therefore directly linked to the level of uncertainty in the observed data.  While the noise term $e_j$ in \eqref{eq:obs_model} represents observation error, generally due to error in measurement or data collection, we can interpret this term as also including modeling error relating to uncertainty in the observation function.  Error due to misfit between the observation function and the data is sometimes referred to as representation error or observation-operator error \cite{Janjic2018, vanLee2014}.  More specifically, we assume that the observation noise covariance matrix $\mathsf{D}$ is the sum of two error covariances, $\mathsf{D} = \mathsf{D}_m + \mathsf{D}_r$, where $\mathsf{D}_m$ represents the error covariance due to measurement noise and $\mathsf{D}_r$ represents the error covariance due to representation error.

\begin{figure}[tb!]
\centering
\includegraphics[width = \textwidth]{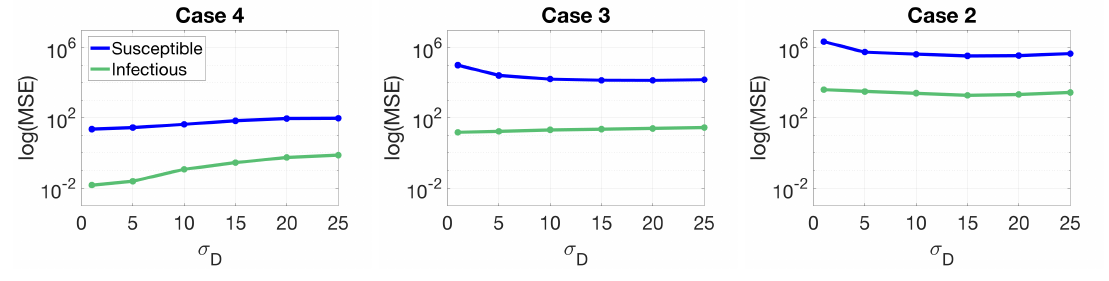}
\caption{
MSE results (in log-scale) averaged over 10 simulations each for increasing values of $\sigma_D$ when using the observation functions in Case 4 (left), Case 3 (middle), and Case 2 (right) for state estimation with known parameter values.}
\label{fig:incD_MSE}
\end{figure}

\begin{figure}[h!]
\centering
\includegraphics[width = \textwidth]{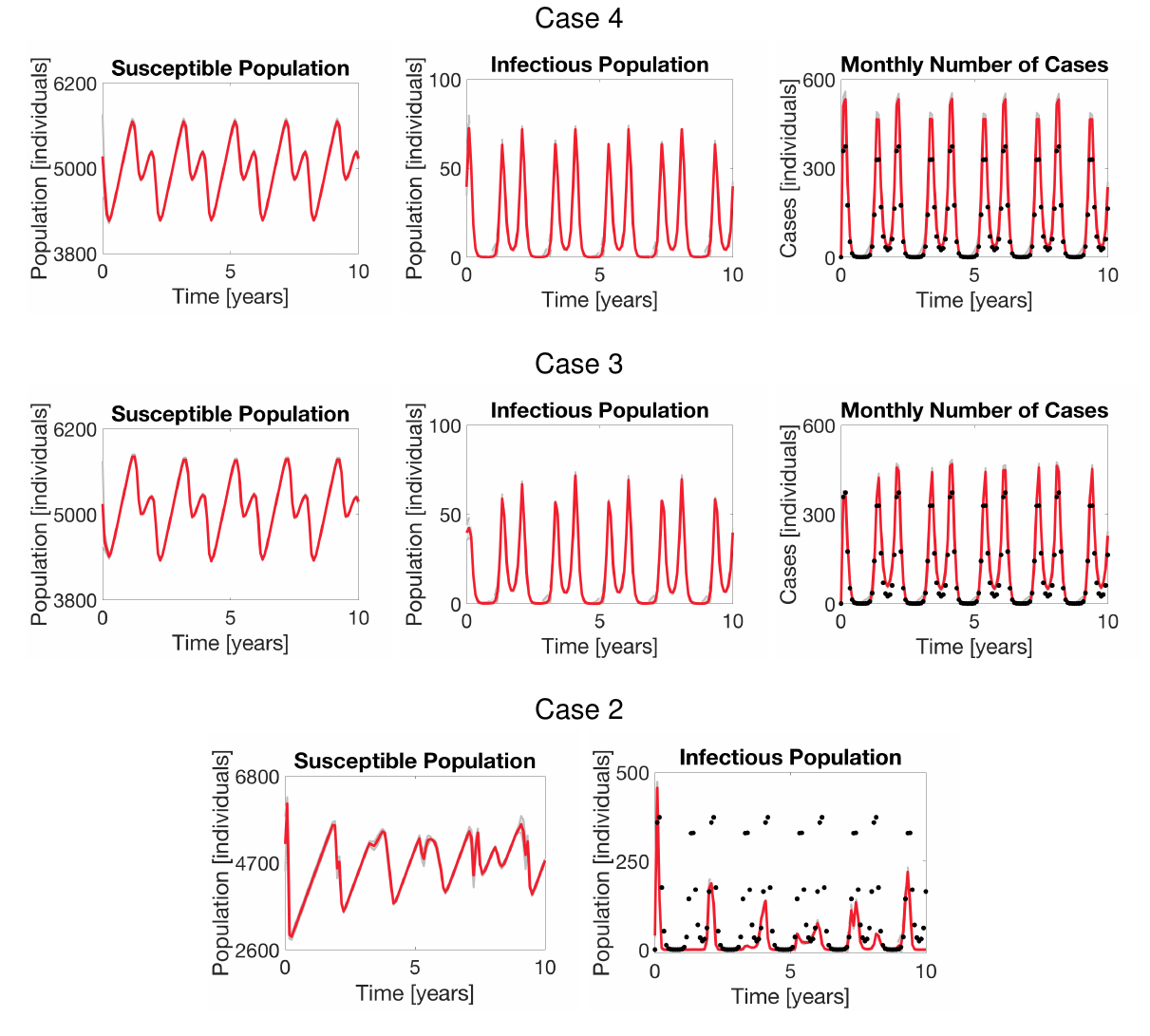}
\caption{Results of the EnKF for state estimation with known parameters for increased $\sigma_D=10$ when employing (from top to bottom): the Case 4 observation function \eqref{eq:case4}; the Case 3 observation function \eqref{eq:case3}; and the Case 2 observation function \eqref{eq:case2}.  In each plot, the EnKF mean is shown in solid red, the $\pm2$ standard deviation curves around the mean are filled in gray, and the observed data are plotted in black markers.   }
\label{fig:incD_og}
\end{figure}

Letting $\mathsf{D} = \sigma_D^2$ with assumed constant variance $\sigma_D^2$, we analyze the effects of increasing the standard deviation $\sigma_{D}$ of the perceived noise in the data to account for additional uncertainty in the observation function selection due to possible representation error.  Figure~\ref{fig:incD_MSE} shows the MSE results (in log-scale) averaged over 10 simulations each for increasing values of $\sigma_D$ (namely, $\sigma_D = 1$, 5, 10, 15, 20, and 25) when using the observation functions in Cases 4, 3, and 2 for state estimation with known parameter values.  Figure~\ref{fig:incD_og} shows state estimation results when using the Case 4, 3, and 2 observation functions, respectively, with $\sigma_{D}=10$ (i.e., $10\times$ the value used in the comparable numerical experiments in Section~\ref{subsec:state_est}).  

As the MSE plots in Figure~\ref{fig:incD_MSE} suggest, increasing $\sigma_D$ when using the true observation function in Case 4 leads to an increase in the MSE for both $S(t)$ and $I(t)$.  Even so, the resulting state estimates retain the shape and magnitude of the true model states with slightly larger uncertainty bounds, as illustrated in Figure~\ref{fig:incD_og}.  For Cases 3 and 2, however, increasing $\sigma_D$ results in an order of magnitude decrease in the MSE for $S(t)$ that remains fairly consistent for $\sigma_D = 10, 15, 20$ and $25$, while maintaining the same order of magnitude in the MSE for $I(t)$.  Compared to the results in Figure~\ref{fig:og} for Case 3 with $\sigma_D = 1$, the results in Figure~\ref{fig:incD_og} demonstrate that increasing $\sigma_D$ (to 10 in this case) allows the filter to better accommodate for the lack of reporting probability in the Case 3 observation function.  The results for Case 2 show that while the magnitude of the $S(t)$ estimate improves compared to the previous results in Figure~\ref{fig:og}, increasing $\sigma_D$ in this case does not improve the overall shape of the state estimates, still notably overestimating the infectious population in several peaking years.  While not shown, similar results hold for Case 1 as for Case 2.  

\begin{figure}[h]
\centering
\includegraphics[width = \textwidth]{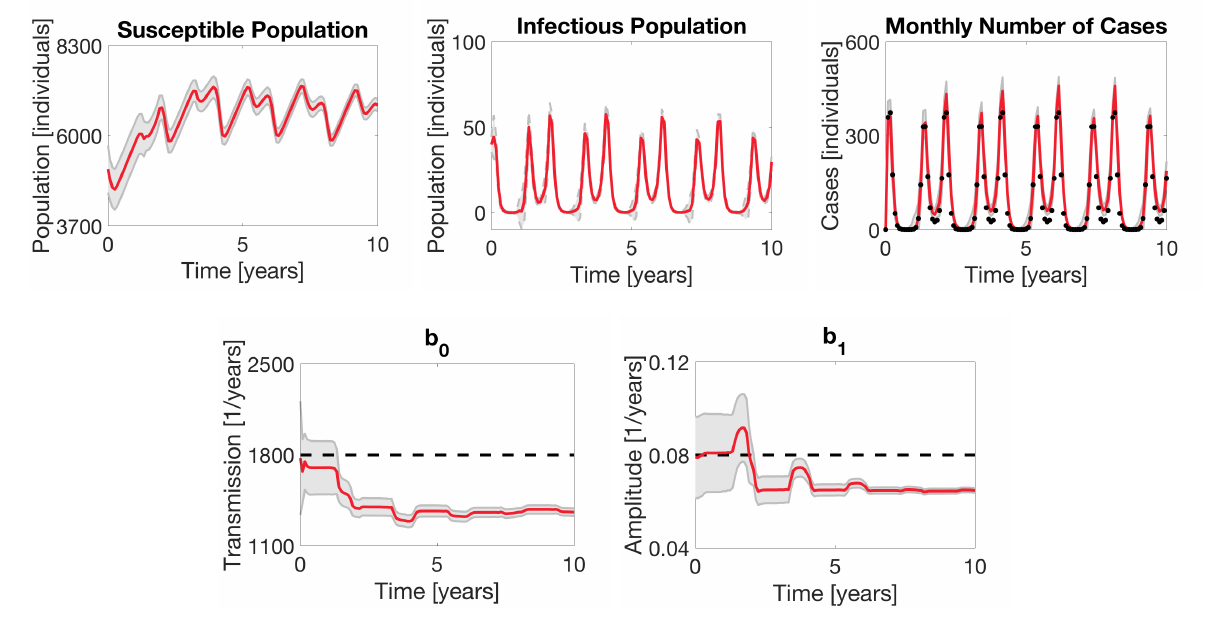}
\caption{
Results of the augmented EnKF for combined state and constant parameter estimation with increased $\sigma_D=25$ when employing the Case 3 observation function \eqref{eq:case3}.  In each plot, the EnKF mean is shown in solid red, the $\pm2$ standard deviation curves around the mean are filled in gray, and the observed data are plotted in black markers.  The true parameter values for $b_0$ and $b_1$ are plotted in dashed black.  Increasing $\sigma_D$ improves convergence in the parameter estimates.
}
\label{fig:incD_case3_aug}
\end{figure}

A similar procedure can be applied to analyze the effects of increasing $\sigma_D$ in the augmented EnKF for constant and time-varying parameter estimation.  For constant parameter estimation, increasing $\sigma_D$ when using the Case 4 observation function slightly increases the uncertainty bounds but still results in parameter estimates for both $b_0$ and $b_1$ that well converge to their true values.  When using the Case 3 observation function, increasing $\sigma_D$ can help to improve the convergence of the resulting parameter estimates, as illustrated in Figure~\ref{fig:incD_case3_aug} with $\sigma_D = 25$.  Compared to the results in Figure~\ref{fig:case3_aug} with $\sigma_D = 1$, while both parameters are still underestimated, here we see the estimates remaining within closer ranges to the true parameter values, especially for $b_0$.  The corresponding estimate of $S(t)$ is also noticeably improved, with an MSE value of $2.39 \times 10^6$.  For time-varying parameter estimation, increasing $\sigma_D$ does not improve the estimates of $\beta(t)$ significantly for any of the observation functions considered.

%%%% FIGURE
%%%%

%%%% FIGURE
%%%%

%%%% FIGURE
%%%%

%%%%%%%%%%%%%%%%%%%%%%%%%%%%%%%%%%%%%%%%%%%%%%%%%%%%%
\section{Discussion} \label{ch:diss}

In this work, we presented a novel computational analysis demonstrating the important role of observation function selection in using ensemble Kalman filtering for state and parameter estimation in the setting of epidemic modeling.  Using a standard SIR model to simulate the forward dynamics, we considered four different observation functions of varying levels of complexity (representing models for prevalence data to under-reported incidence data) within three EnKF frameworks: state estimation with known parameters; combined state and constant parameter estimation; and combined state and time-varying parameter estimation.  With synthetic data generated using the under-reported incidence observation function in Case 4, we showed how incorrect observation modeling assumptions misinterpreting the data (by mistaking incidence data for prevalence data, or by neglecting the effects of under-reporting) result in correspondingly inaccurate estimates of the SIR model states and unknown parameters in each of the scenarios considered. 

For the state estimation with known parameters in Figure~\ref{fig:og}, when using true the observation function in Case 4, the filter was able to estimate the true model states with a high accuracy.  When using the observation function in Case 3, the absence of the reporting probability $\rho$ mainly affected the magnitude of the results, initially underestimating $I(t)$ with some larger peaks occurring bi-annually in later years and consistently overestimating $S(t)$.  In the estimated monthly number of cases, the filter fits to the observed data points and does not account for under-reporting due to the lack of the reporting probability parameter in the Case 3 observation function.  The results when using the observation functions in Cases 2 and 1 further degrade: These functions interpret the under-reported incidence data as direct observations of the infectious population (Case 1) or a proportion of the infectious population (Case 2), leading to issues in both the magnitude and shape of the corresponding state estimates.  

Combined state and parameter estimation increases the complexity of the problem, since the parameters relating to disease transmission are assumed to be unknown and therefore must be estimated along with the model states.  For constant parameter estimation, the filter converged to the true values of $b_{0}$ and $b_{1}$ after approximately 2 years of data when using the Case 4 observation function, as shown in Figure~\ref{fig:case4_aug}.  However, when using the function in Case 3, the filter had significant difficulty estimating the parameters: While the estimate for $b_{1}$ stayed within the same general value region, neither estimate well converged and the estimate for $b_{0}$ diverged towards about half of its true value, as shown in Figure~\ref{fig:case3_aug}.  The effects are also clearly seen in the estimate for $S(t)$, which no longer tracks the true shape and increases well outside of the original range of values.  As with the state estimation, parameter estimation results further degraded when using the observation functions in Cases 1 and 2, with the parameter estimates diverging to inaccurate values.  For additional analysis of these effects, we note that the posterior EnKF mean and variance estimates for the constant parameters could be compared to the posterior samples obtained from a Markov Chain Monte Carlo (MCMC)-type algorithm; see, e.g., \cite{MCMC, DRAM}.

For time-varying parameter estimation, tracking the full transmission function $\beta(t)$ without assuming a known form was challenging even when using the true observation function in Case 4, as shown in Figure~\ref{fig:case4_pd}.  While the filter mean estimate generally follows the shape and the true transmission function is fully captured within relatively small uncertainty bounds, the filter estimate is less smooth and does not maintain the periodicity of the true function.  However, Figure~\ref{fig:case3_pd} shows that when using the observation function in Case 3, the filter mean estimate drifts away from the true transmission function, thereby underestimating transmission and overestimating the susceptible population.  Parameter tracking using the functions in Cases 1 and 2 showed similar tendencies to diverge and inaccurately estimate the time-varying transmission parameter.

Numerical experiments also considered the effects of increasing the standard deviation $\sigma_D$ of the observation noise in the filter to account for potential representation error due to uncertainty in the choice of observation function.  The observation functions considered in this work range from linear misfit (between Case 3 and Case 4) to more substantial nonlinear discrepancies (between Cases 1 and 2 and Case 4).  As illustrated in Figure~\ref{fig:incD_og} for state estimation, when using the true observation function in Case 4, increasing $\sigma_D$ caused a small increase in corresponding MSE values for $S(t)$ and $I(t)$ but overall did not negatively impact the estimation.  Results in Figures~\ref{fig:incD_og} and \ref{fig:incD_case3_aug} further show that increasing $\sigma_D$ helped to improve state estimation and constant parameter estimation results when using the observation function in Case 3, which differs from the true observation function by a constant.  However, while the results for Case 2 show some improvement in magnitude for the estimated $S(t)$, the overall shape of the true function was not captured.  These results suggest that increasing the observation noise covariance may help to offset the effects of a missing constant or linear difference in the observation function, but it is unable to fully account for more significant modeling discrepancies.

Further, the results in Figure~\ref{fig:incD_MSE} suggest that continuing to increase $\sigma_D$ does not result in better estimates after a certain point, with MSE values for Cases 2 and 3 leveling off after $\sigma_D=10$ in the state estimation considered.  In certain cases, continued increase of $\sigma_D$ may also lead to numerical integration issues.  Future work aims to incorporate systematic mechanisms within the EnKF framework to automatically adjust $\sigma_D$ to account for suboptimal observation function selection, with the goal of improving filter estimates while avoiding overinflation.  Recently proposed alternative methods for addressing representation error in data assimilation include: using the ensemble covariance to update the observation noise covariance \cite{Satterfield2017}; applying an iterative scheme to adjust the observation function with nearest neighbors information \cite{chaos}; and using a secondary filter with tools from machine learning to estimate the representation error \cite{Berry2017}.  

With the increased use of ensemble Kalman-type filtering for parameter estimation and model forecasting in epidemic applications, our results serve to illustrate the potential limitations in filter performance when the observation function misinterprets or inaccurately represents the available data.  This is especially important to acknowledge when using real epidemiological data to make forecast predictions, as inaccurate state and parameter estimates lead to diminished accuracy in corresponding forecasts.  While we have focused on variants of the EnKF, we note that other data assimilation algorithms are applicable in this setting, e.g., \cite{Gilks2001, Katzfuss2020, Arnold2013, Nadler2020, Rhodes2009}.  In future work, we aim to address observation function selection in the context of particle filtering algorithms \cite{tutorial, SMCbook} and related techniques, as well as deterministic optimization algorithms \cite{Johnson1992, Banks2014} for parameter estimation in epidemiological applications.  Future work also includes development of modified algorithms to systematically address suboptimal observation function selection, with the goal of adjusting for potential misinterpretation of observed data on the fly as the assimilation proceeds.

%%%%%%%%%%%%%%%%%%%%%%%%%%%%%%%%%%%%%%%%%%%%%%%%%%%%%
\section{Summary and Conclusions}\label{ch:conclusion}

The observation function plays a critical role in connecting the observed data with the forward model variables in the ensemble Kalman filtering framework.  In this work, we present a novel simulation study analyzing the effects of observation function selection when using the EnKF for state and parameter estimation in the setting of epidemic modeling, where key differences in observed data have led to the use of different observation functions in the literature.  In examining the use of four epidemiologically-inspired observation functions of different forms in connection with the classic SIR model, we show how incorrect observation modeling assumptions (i.e., assimilating incidence data with a prevalence model, or neglecting the effects of under-reporting) can lead to inaccurate filtering estimates and correspondingly inaccurate forecast predictions.  Our results demonstrate the importance of choosing an observation function that well interprets the given data on the corresponding EnKF state and parameter estimates for both constant and time-varying parameters, especially as the problem becomes more difficult by including additional unknowns.  Numerical experiments further illustrate how increasing the observation noise covariance can help to account for representation error in the selected observation function in cases with linear misfit.

%%%%%%%%%%%%%%%%%%%%%%%%%%%%%%%%%%%%%%%%%%%%%%%%%%%%%%
\section*{Acknowledgments}

The authors would like to thank the Department of Mathematical Sciences at WPI, with a special thanks to Suzanne Weekes, Sarah Olson, Michael Yereniuk, and Caroline Johnston, for encouraging this research.\\[0.2cm]

\noindent \textbf{Funding:} This work was partially supported by the Henry Luce Foundation under grant number 9136 to WPI (Clare Boothe Luce Research Scholarship to L. Mitchell) and the National Science Foundation under grant number NSF/DMS-1819203 (PI A. Arnold).\\[0.2cm]

\noindent \textbf{Declaration of Competing Interest:} None.\\[0.1cm]

%%%%%%%%%%%%%%%%%%%%%%%%%%%%%%%%%%%%%%%%%%%%%%%%%%%%%%
\begin{appendices}

\renewcommand\thefigure{\thesection.\arabic{figure}} 
\setcounter{figure}{0}

%%%%%%%%

% Appendix A
\section{Observation Noise Models}
\label{Appendix:Noise}

The synthetic data generated in Section~\ref{subsec:data_gen} assumes an additive noise model of the form
\begin{equation}
y_{j}=g(x_{j},\theta) + e_{j}, \hspace{.2in} j=1,\dots,T
\end{equation}
where $y_{j}\in\R$ is the observed data at time $t_j$, $g(x_{j},\theta)$ is the observation function relating the model states $x_{j}=x(t_j)$ and parameters $\theta$, and $e_{j}\in\R$ is the observation error, normally distributed with mean zero and variance $\sigma_\text{noise}^2$. 

For the noise level in this work, the additive noise model generally yields non-negative observations of the under-reported monthly incidence data.  However, in order to guarantee non-negative observations and add noise proportional to the number of cases, we could consider instead a multiplicative noise model of the form
\begin{equation}
y_{j}=e_{j}g(x_{j},\theta), \hspace{.2in} j=1,\dots,T
\end{equation}
which is additive on the log-scale, with
\begin{equation}
\log(y_{j})=\log(g(x_{j},\theta)) + \log(e_{j}), \hspace{.2in} j=1,\dots,T
\end{equation}
where $\tilde{e}_j = \log(e_{j})$ is normally distributed with mean zero and variance $\tilde{\sigma}_\text{noise}^2$.

\begin{figure}[tb!]
\centering
\includegraphics[width = \textwidth]{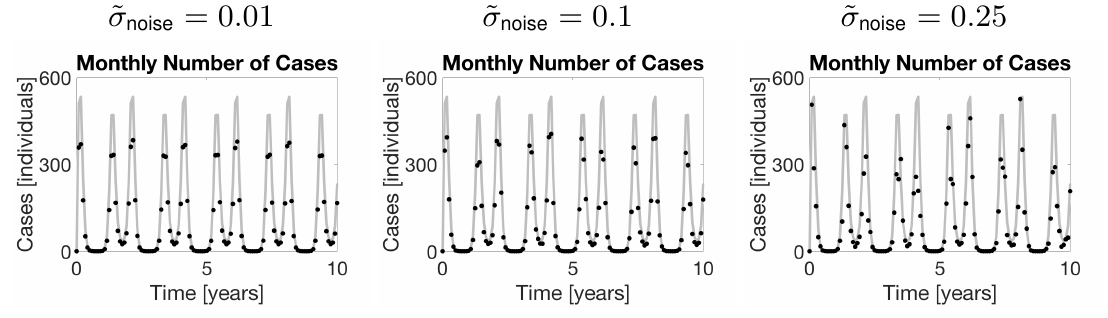}
\caption{
Synthetic data generated from the SIR model \eqref{eq:SIR1}--\eqref{eq:SIR2} with parameters given in Table~\ref{Tab:Params} using the Case 4 observation function \eqref{eq:case4} at 70\% reporting.  Here the observations are corrupted using the multiplicative noise model with different values of the standard deviation $\tilde{\sigma}_\text{noise}$.  In each plot, the true number of monthly cases are shown in gray, and the observed data are displayed as black markers. generated  and corrupted with Gaussian noise. 
}
\label{fig:noise_models}
\end{figure}

Figure~\ref{fig:noise_models} shows three sets of simulated data, generated as in Section~\ref{subsec:data_gen} but corrupted instead using the multiplicative noise model with $\tilde{\sigma}_\text{noise} = 0.01$, $\tilde{\sigma}_\text{noise} = 0.1$, and $\tilde{\sigma}_\text{noise} = 0.25$, respectively.  When compared to the data in Figure~\ref{fig:synth_data}, which was corrupted using the additive noise model with $\sigma_\text{noise}=0.1$, we note that the multiplicative noise model yields similar results for $\tilde{\sigma}_\text{noise} = 0.01$ but the error is visibly larger for $\tilde{\sigma}_\text{noise} = 0.25$ with respect to the higher case numbers.

%%%% FIGURE
%%%%

%%%%%%%%%%%%%%%%%%%%%%%

% Appendix B
\section{Filter Consistency Check}
\label{Appendix:Consistency}

As discussed in \cite{enkftheory}, it is possible to check the consistency of the EnKF by comparing the statistics computed from the filter innovation sequence with the sum of the observation noise and forecast error covariances.  The innovation sequence is given by
\begin{equation}
\nu_{j} = \frac{1}{N} \ds\sum_{n=1}^{N} (y_{j}^{n} - \hat{y}_{j}^n) (y_{j}^{n} - \hat{y}_{j}^n)^\mathsf{T}, \hspace{.2in} j=1,\dots,M
\end{equation}
where $y_{j}^{n}$ is an artificial observation as defined in \eqref{eq:artificial_obs} and $\hat{y}^n_{j}$ is the observation prediction computed using the observation function in \eqref{eq:enkf_obspred}.  The sum of the observation noise covariance $\mathsf{D}$ and the forecast error covariance $\mathsf{\Phi}^{\hat{y}\hat{y}}_{j}$ appears in the Kalman gain \eqref{eq:enkf_kg}. 

Since the observations in this work (i.e., the number of observed cases each month) are scalar quantities, it follows that the innovation terms $\nu_j$, the observation noise covariance $\mathsf{D} = \sigma_D^2$, and forecast errors $\mathsf{\Phi}^{\hat{y}\hat{y}}_{j} = \phi^{\hat{y}\hat{y}}_{j}$ are all scalars, and we can compare these values as a ratio averaged over time, such that
\begin{equation}\label{eq:consist_approx}
\gamma = \frac{1}{M} \ds\sum_{j=1}^{M} \frac{\sigma_D^2 + \phi^{\hat{y}\hat{y}}_{j}}{\nu_{j}}
\end{equation}
gives an approximate measure of consistency for the filter, where $M$ is the number of observations.  

\begin{figure}[tb!]
\centering
\includegraphics[width = 0.5\textwidth]{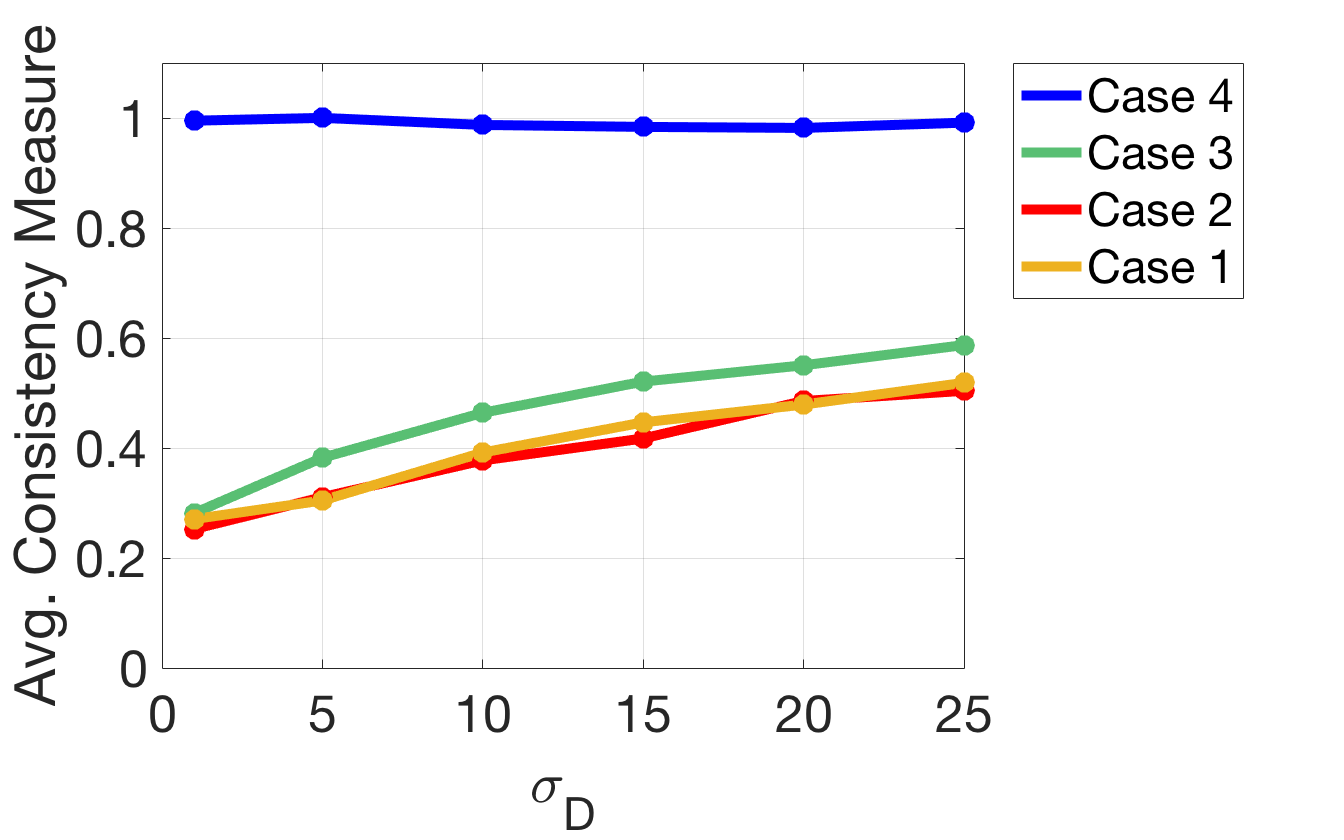}
\caption{
Approximate consistency measure $\gamma$ in \eqref{eq:consist_approx} averaged over 5 runs of the EnKF for state estimation with known parameter values, using the synthetic data generated in Section~\ref{subsec:data_gen} and each of the four observation functions for increasing values of the observation noise standard deviation $\sigma_D = 1$, 5, 10, 15, 20, and 25.
}
\label{fig:consistency}
\end{figure}

Figure~\ref{fig:consistency} shows the approximate consistency measure $\gamma$ in \eqref{eq:consist_approx} averaged over 5 runs of the EnKF for state estimation with known parameter values, using the synthetic data generated in Section~\ref{subsec:data_gen} and each of the four observation functions for increasing values of the observation noise standard deviation $\sigma_D = 1$, 5, 10, 15, 20, and 25.  When using the Case 4 observation function (which is the true observation function used in generating the data), note that the approximate consistency measure is around 1 for all values of $\sigma_D$, suggesting fairly reasonable predicted error statistics.  When using the other three observation functions, the approximate consistency measure is lower, ranging between 0.2 and 0.6 but increasing with increased $\sigma_D$ in each case.

%%%% FIGURE
%%%%

\end{appendices}

%%%%%%%%%%%%%%%%%%%%%%%%%%%%%%%%%%%%%%%%%%%%%%%%%%%%%%
%%%%%%%%%%%%%%%%%%%%%%%%%%%%%%%%%%%%%%%%%%%%%%%%%%%%%%

% Bibliography

\bibliography{bibl}{}

\end{document}